\documentclass{ws-ijmpa}
\usepackage[super,compress]{cite}
\usepackage{amsmath}
\usepackage{amsfonts}
\usepackage{amssymb}
\usepackage{hyperref}
\usepackage{float}
\usepackage[all]{hypcap}
\usepackage{paralist}
\usepackage{multirow}                     
\usepackage{dcolumn}
\usepackage{paralist}
\usepackage[T1]{fontenc} 
\usepackage[textsize=scriptsize,backgroundcolor=red!70,linecolor=red]{todonotes}
\usepackage{graphicx}
\usepackage{subfigure}
\newcommand{\dcp}{\ensuremath{\delta_{\text{CP}}}}
\newcommand{\nue}{\ensuremath{\nu_e}}
\newcommand{\numu}{\ensuremath{\nu_\mu}}

\newcommand{\anue}{\ensuremath{\bar{\nu}_e}}
\newcommand{\anumu}{\ensuremath{\bar{\nu}_\mu}}

\newcommand{\chisq}{\ensuremath{\chi^2}}
\newcommand{\anti}[1]{\ensuremath{\bar{#1}}}
\newcommand{\liar}{LAr}
\newcommand{\ie}{\textit{i.e.}}
\newcommand{\eg}{\textit{e.g.}}

\newcommand{\viz}{\textit{viz.}}

\newcommand{\dmsqb}{\ensuremath{\Delta m^{2}_{21}}}
\newcommand{\upmns}{\ensuremath{U_\text{PMNS}}}

\newcommand{\pmue}{\ensuremath{P_{\mu e}}}
\newcommand{\pmumu}{\ensuremath{P_{\mu \mu}}}
\newcommand{\dchisq}{\ensuremath{\Delta \chisq}}
\newcommand{\pimh}{\ensuremath{\Pi^\text{MH}}}

\newcommand{\nova}{\ensuremath{\text{NO}\nu\text{A}}}
\def\be{\begin{equation}}
\def\ee{\end{equation}}
\def\bea{\begin{eqnarray}}
\def\eea{\end{eqnarray}}
\def\gsim{\ \rlap{\raise 2pt\hbox{$>$}}{\lower 2pt \hbox{$\sim$}}\ }
\def\lsim{\ \rlap{\raise 2pt\hbox{$<$}}{\lower 2pt \hbox{$\sim$}}\ }
\def\dslash{\kern-4pt \not{\hbox{\kern-2pt $\partial$}}}
\def\pslash{\not{\hbox{\kern-2pt p}}}

\begin{document}
\DeclareGraphicsExtensions{.eps,.ps, .pdf}
\title{Optimal configurations of the Deep Underground Neutrino Experiment}
\author{Vernon Barger\footnote{barger@pheno.wisc.edu}}
\address{Department of Physics, University of Wisconsin, Madison,
 WI 53706, USA}
\author{Atri Bhattacharya\footnote{atrib@email.arizona.edu}}
\address{Department of Physics, University of Arizona, Tucson, AZ 85721, USA}
\author{Animesh Chatterjee\footnote{animesh.chatterjee@uta.edu}}
\address{University of Texas at Arlington, Arlington, TX 76019, USA}
\author{Raj Gandhi\footnote{nubarnu@gmail.com}}
\address{Harish-Chandra Research Institute, Chhatnag Road, Jhunsi,
Allahabad 211 019, India}
\author{Danny Marfatia\footnote{dmarf8@hawaii.edu}}
\address{Department of Physics and Astronomy, University of Hawaii at Manoa,
Honolulu, HI 96822, USA}
\author{Mehedi Masud\footnote{masud@hri.res.in}}
\address{Harish-Chandra Research Institute, Chhatnag Road, Jhunsi,
Allahabad 211 019, India}
\maketitle
\begin{abstract}
%\abstract{
We perform a comprehensive study of the ability of the Deep Underground Neutrino Experiment (DUNE) to
answer outstanding questions in the neutrino sector. We consider the sensitivities to the mass hierarchy, the octant of $\theta_{23}$ and to CP violation using data from beam and atmospheric neutrinos.
We evaluate the dependencies on the precision with which $\theta_{13}$ will be measured by reactor experiments, on the detector size, beam power and exposure time, on detector magnetization, and on the systematic uncertainties achievable with and without a near detector. We find that a 35 kt far detector in DUNE with a near detector will resolve the eight-fold degeneracy that is intrinsic to long baseline experiments and will meet the primary goals of oscillation physics that it is designed for.
\end{abstract}
%\maketitle
\section{Introduction}
\label{sec:intro}
Neutrino oscillations have by now been
conclusively established by several pioneering experiments.
It is now understood that the mixing between the three 
neutrino flavors is governed by the so-called PMNS mixing matrix,
\begin{equation}\label{eqn:UPMNS}
	\upmns = 
	\begin{pmatrix}
    c_{12} c_{13}
    & s_{12} c_{13}
    & s_{13} e^{-i\delta_\text{CP}}
    \\
    - s_{12} c_{23} - c_{12} s_{13} s_{23} e^{i\delta_\text{CP}}
    & \hphantom{+} c_{12} c_{23} - s_{12} s_{13} s_{23}
    e^{i\delta_\text{CP}}
    & c_{13} s_{23} \hspace*{5.5mm}
    \\
    \hphantom{+} s_{12} s_{23} - c_{12} s_{13} c_{23} e^{i\delta_\text{CP}}
    & - c_{12} s_{23} - s_{12} s_{13} c_{23} e^{i\delta_\text{CP}}
    & c_{13} c_{23}	\hspace*{5.5mm}	
	\end{pmatrix}\,, 
\end{equation}
and the mass-squared differences: $\Delta m^2_{31}=m_3^2-m_1^2$ and \dmsqb. 
Here, $ c_{ij} $ and $ s_{ij} $ are $ \cos\theta_{ij} $ and $ \sin\theta_{ij} $
respectively, for the three mixing angles 
$ \theta_{12} $, $ \theta_{23} $ and $ \theta_{13} $, and \dcp\ is a
(Dirac) CP phase.
While solar and atmospheric neutrino experiments have determined 
 the first two mixing angles quite precisely, reactor experiments
have made remarkable progress in determining
$ \theta_{13} $~\cite{db}.
(See Table~\ref{tab:bestfit} for the values of the oscillation parameters used in our work.)

\begin{table*}[t]
\tbl{Best fit values of the oscillation parameters \cite{Fogli}.}
{
\centering
\begin{tabular}{|c|c|}
\hline 
\textbf{Parameter}    & \textbf{Best fit value}  \\ \hline
$\sin^{2}\theta_{12}$ & 0.307 \\ \hline
$\sin^{2}\theta_{13}$ & 0.0241 \\ \hline
$\sin^{2}\theta_{23}$ (lower octant) &  0.427\\ \hline
$\sin^{2}\theta_{23}$ (higher octant) &  0.613\\ \hline
$\Delta m_{21}^{2}$ & $7.54 \times 10^{-5} \text{ eV}^{2}$ \\ \hline
$|\Delta m_{31}^{2}|$ & $2.43 \times 10^{-3} \text{ eV}^{2}$ \\ \hline
$\delta_{CP}$ & 0 \\
\hline
\end{tabular} \label{tab:bestfit}}
\end{table*}

Now that $\theta_{13}$ has been conclusively shown to be non-zero and not too small~\cite{db,osc}, the focus of neutrino
oscillation experiments has shifted to the measurement of $\delta_{CP}$ that determines 
whether or not oscillating neutrinos violate CP. A second important unanswered question for model building is whether the mass hierarchy is {\it normal} with $\Delta m^2_{31}>0$, or {\it inverted} with $\Delta m^2_{31}<0$. 
Finally, the question of whether $\theta_{23}$ is larger or smaller than $\pi/4$ bears on models based on
lepton symmetries.

An effort towards resolving the above issues and thereby taking us a step closer to completing our knowledge of the neutrino mass matrix, is the Deep Underground Neutrino Experiment (DUNE).\footnote{The inputs we use, and the corresponding references, pertain  to the
erstwhile Long Baseline Neutrino Experiment (LBNE)\cite{LBNE, LBNE-interim}, which  has  undergone  a new
phase of
internationalisation and expansion. This has led to a change in the name of the
experiment, to DUNE. Nonetheless, it is expected that the configuration we
assume here vis a
vis fluxes, baseline and energies will remain largely
intact.} 
DUNE will employ a large liquid argon far detector (FD). It is expected to be placed underground in the Homestake mine at a distance of 1300~km from Fermilab, from which a neutrino beam will be directed towards the detector. Large-mass Liquid Argon 
Time Projection Chambers (LAr-TPCs) have 
unprecedented capabilities for the detection of neutrino 
interactions  due to precise and 
sensitive spatial and calorimetric resolution. However, the final configuration of the experiment is still under discussion~\cite{reconfig,Barger:2013}. The sensitivity of DUNE to the mass hierarchy, to CP violation and to the octant of $\theta_{23}$ depends on, among other things, how well other oscillation parameters are known, on the amount of data that can be taken in a reasonable time frame, on the systematic uncertainties that compromise the data, and on the charge discrimination capability of the detector. Our goal (in this extension of our previous work~\cite{Barger:2013}) is to study how these factors affect DUNE in various experimental configurations. Other recent studies of some of the physics capabilities of DUNE can be found in Ref.~\cite{Bass:2013taa} .

\subsection{Objectives}
\label{sec:obj}

The various considerations of our work are motivated by possible configurations for DUNE in the initial
phase of its program.
The initial stage of DUNE will, at the very least, permit the construction of
an unmagnetized 10~kt FD underground. 
Several improvements upon this basic configuration are under consideration.
These might include 
\begin{itemize}
	\item upgrading the FD to 35~kt for improved statistics,  
	\item having a precision near detector (ND) for better calibration of the initial flux
	      and reducing the involved systematic uncertainties,
	\item  magnetizing the FD to make it possible to
	      distinguish between particles and antiparticles in the atmospheric neutrino
	      flux.\footnote{The beam experiment would have the neutrino and
	      antineutrino runs happen asynchronously. Thus, the events from the two would be
	      naturally distinguished and magnetization of the FD would not affect its results.}
\end{itemize}
It must be noted that some of the above upgrades would also have supplementary benefits
--- an ND, for example, will also allow precision studies of the involved
neutrino nucleon cross sections, thereby reducing present uncertainties.

Since it might not be feasible to combine all of the above upgrades into an initial DUNE configuration, we evaluate which combination would be most beneficial as far as the physics of neutrino oscillations is concerned.
Specifically,  we study the following experimental configurations:
\begin{enumerate}
	\item A beam experiment with and without an ND.
	\item An atmospheric neutrino experiment.
	\item An experiment with and without an ND that combines beam and atmospheric neutrino data collected at the FD.
      \item A global configuration that combines DUNE data (with and without ND) with NO$\nu$A~\cite{nova} and T2K~\cite{t2k} data.
\end{enumerate}
Our study will highlight the benefits of
\begin{enumerate}

	\item building a larger 35 kt FD as opposed to a 10 kt detector,
         \item higher exposure (kt-MW-yr)~\cite{Barger:2006kp},
	\item magnetizing the FD versus having an unmagnetized detector volume,
	\item utilizing atmospheric neutrinos,
         \item precision $\theta_{13}$ measurements,

\end{enumerate}
Throughout, we estimate how data from NO$\nu$A
and T2K improve DUNE sensitivities; for previous discussions see Refs.~\cite{LBNE, Agarwalla:2012bv} . 
All simulations for the beam experiments have been done numerically with the GLoBES 
software~\cite{globes}.

\section{Experimental specifications and analysis methodology}
\label{sec:Det}

We consider neutrinos resulting from a 120~GeV proton beam from Fermilab with a beam power of 700 kW and an uptime of $1.65\times 10^7$ seconds per year (or equivalently $6 \times 10^{20}$ protons on target (POT) per year) incident at a \liar\ FD at a baseline of 1300~km; an upgrade to a 2.3~MW beam is a possibility. As noted in Ref.~\cite{pref} , the physics performance is roughly equivalent for proton beam energies that exceed 60~GeV.  We simulate events at the FD using the GLoBES software~\cite{globes} and the fluxes used by the DUNE collaboration.

If the FD is placed underground, it is also sensitive to atmospheric neutrinos.
We simulate atmospheric neutrino events (as described in Appendix~\ref{sec:num-atm}) and both \nue\ appearance and \numu\ disappearance channel events from the beam in the neutrino and antineutrino modes with an event reconstruction efficiency of 85\%. In our simulation of the DUNE beam experiment we employ the signal and background systematics for \nue\ appearance and \numu\ disappearance channels from Refs.~\cite{LBNE-interim,sanjib} ; see Table~\ref{tab:syst}.{\footnote{Our ND analysis represents the most obvious benefit that the beam experiment will reap with an ND, \viz, improvement in systematics for the signal and background events.
In addition, an ND will also  improve our understanding of the fluxes and cross sections, thereby reducing shape-related uncertainties in the analysis.
We do not attempt an exploration of this facet of the ND because the exact nature of the improvement would depend to a large extent on the specifics of the ND, which for the DUNE is yet in the planning stage.
Our ND analysis represents a worst-case scenario for improvement in the systematics.
}
 For the energy resolutions, we have used the method of bin-based automatic energy smearing with $\frac{\sigma}{E} = \frac{0.20}{\sqrt{E}}$ for \numu\ events and $\frac{\sigma}{E} = \frac{0.15}{E}$ for \nue\ events; see the appendix of Ref.~\cite{LBNE-interim}. An alternative is to use migration matrices~\cite{LBNE-interim, Agarwalla:2012bv}. For NO$\nu$A~\cite{nova} and T2K~\cite{t2k}, the relevant parameters are given in Tables~\ref{tab:syst-nova} and~\ref{tab:syst-t2k}, respectively. An up-to-date description of the NO$\nu$A and T2K experiments can be found in Ref.~\cite{Agarwalla:2012bv} .  

\begin{table}[t]
\tbl{Systematic uncertainties for signal and background channels in 
DUNE\cite{LBNE-interim,sanjib}.}
{
\centering
\begin{tabular}{|c|c|c|c|}
\hline
\multirow{2}{*}{Detector configuration} & \multicolumn{2}{c|}{Systematics} \\ \cline{2-3}
                                        & Signal       & Background \\ \hline
\multirow{2}{*}{With ND	}               & \nue  : 1\%  & \nue  : 1\% \\
                                        & \numu : 1\%  & \numu : 5\% \\ \hline 
\multirow{2}{*}{Without ND}             & \nue  : 5\%  & \nue  : 10\% \\
                                        & \numu : 5\%  & \numu : 45\% \\
\hline
\end{tabular} \label{tab:syst}}
\end{table}

\begin{table}[t]
\tbl{Systematic uncertainties for NO$\nu$A \cite{nova}.}
{
\centering
\begin{tabular}{|c|c|c|c|}
\hline 
\multirow{2}{*}{Detector configuration} & \multicolumn{2}{c|}{Systematics} \\ \cline{2-3}
                                        & Signal & Background \\ \hline
15 kt TASD	                            & \nue : 5\% & \nue : 10\% \\ 
3 yrs. $\nu$ + 3 yrs. $\bar{\nu}$ &  & \\ 
$6 \times 10^{20}$ POT/yr	& \numu : 2\% & \numu : 10\% \\
with a 700 kW beam &  & \\
\hline
\end{tabular} \label{tab:syst-nova}}
\end{table}

\begin{table}[t]
\tbl{Systematic uncertainties for T2K \cite{t2k}.}
{
\centering
\begin{tabular}{|c|c|c|c|}
\hline 
\multirow{2}{*}{Detector configuration} & \multicolumn{2}{c|}{Systematics} \\ \cline{2-3}
                                        & Signal & Background \\ \hline
22.5 kt water Cherenkov	& \nue : 5\% & \nue : 5\% \\ 
5 yrs. $\nu$  &  & \\ 
$8.3 \times 10^{20}$ POT/yr	& \numu : 5\% & \numu : 5\% \\
with a 770 kW beam &  & \\
% &  & \\
\hline
\end{tabular} \label{tab:syst-t2k}}
\end{table}

Throughout, we assume that placing the detector underground does not result in a significant change of the signal and background analysis, apart from making the detector also sensitive to the atmospheric neutrino events, and thus allowing a combined analysis of the beam and atmospheric data over the duration of the experiment.

We assume that the beam is run in the neutrino mode for a period of five years, and thereafter in the antineutrino mode for five more years.

For the atmospheric neutrino analysis, and consequently the combined beam and atmospheric analysis, it becomes important to consider both a magnetized and an unmagnetized \liar~detector.
In the former case, the detector sensitivity, especially for the resolution of the mass hierarchy, is significantly improved over the latter, due to its ability to distinguish between particles and antiparticles. 
This, however, is only partly applicable to the \nue\ events, because for a 10~kt volume detector, it is difficult to distinguish between the tracks arising from of \nue\ and \anue\ interactions. This difficulty arises because pair-production and bremsstrahlung sets in with increasing energies, and above $\approx 5$~GeV, the detector completely loses its ability to distinguish between these CP conjugate pairs.
On the other hand, due to their tracks being significantly longer, \numu\ and \anumu\ events are clearly distinguishable at all accessible energies.
We implement this in our detector simulation for the atmospheric neutrino and combined analysis.

For our simulation of atmospheric neutrino data, the energy and angular resolutions of the detector are as in Table~\ref{tab:exptparams}~\cite{Barger:2012fx}. The atmospheric fluxes are taken from Ref.~\cite{Honda:2004yz}  , the flux and systematic uncertainties from Ref.~\cite{raj2007} , and
the density profile of the earth from Ref.~\cite{PREM} .
\begin{table}[t]
\tbl{Detector parameters used for the analysis of atmospheric neutrinos \cite{Barger:2012fx}.}
{
\centering
\begin{tabular}{lc}
\hline
\multirow{2}{*}{Rapidity (y)} 
  & 0.45 for $\nu$\\
  & 0.30 for $\bar{\nu}$\\
\hline\noalign{\smallskip}
Energy Resolution ($\Delta E$) & $\sqrt{(0.01)^{2} + (0.15)^{2}/(yE_{\nu}) + (0.03)^{2}}$\\[2pt]
\hline
\multirow{2}{*}{Angular Resolution ($\Delta \theta$)}
  & $3.2^{\circ}$ for $\nu_{\mu}$	\\
  & $2.8^{\circ}$ for $\nu_{e}$ \\
\hline
Detector efficiency ($\mathcal{E}$) & $85\%$ \\
\hline
\end{tabular} \label{tab:exptparams}}
\end{table}

The charge identification capability of the detector is incorporated 
as discussed in Ref.~\cite{Barger:2012fx} . For electron events, we 
conservatively assume a 20$\%$ probability of  charge identification  
in the energy range $1 - 5$~GeV, and no capability for events with energies 
above 5~GeV. Since the muon charge identification capability of a 
LAr-TPC is excellent for energies between 1 and 10~GeV, we have 
assumed it to be 100\%.

For the combined analysis of beam and atmospheric events, we first calculate the \chisq\ separately from the atmospheric analysis (using our code) and the DUNE beam analysis (using GLoBES) for a set of fixed oscillation parameters. After adding these two fixed parameter $\chi^{2}$ values, we marginalize over $\theta_{13}$, $\theta_{23}$, $|\Delta m_{31}^{2}|$ and \dcp\ to get the minimized \chisq; see Appendix~\ref{sec:num-atm}.
The procedure is similar for our combined analysis of DUNE, T2K and \nova\ data.

\section{Mass hierarchy}
\label{sec:MH}

Since \dcp\ will likely remain undetermined by experiments preceeding DUNE, we analyze the sensitivity to the  mass hierarchy as a function of this parameter.
The analysis is carried out by assuming one of the hierarchies to be true and then determining by means of a \chisq\ test, how well the other hierarchy can be excluded.
We marginalize over the present day uncertainties of each of the prior determined parameters.

The \dcp\ dependence of the sensitivity to the mass hierarchy arises through the oscillation probability\cite{degen},

\begin{equation} \label{eq:pme}
	P_{\mu e} = T_{1}^{2} + T_{2}^{2} + 2T_{1}T_{2}\cos(\dcp + \Delta)\,,
\end{equation}
where,
\begin{eqnarray}
	T_{1} &= \alpha \sin2\theta_{12} \cos\theta_{23} \frac{\sin(x \Delta)}{x} \label{eq:pme_t1}\\
	T_{2} &= \sin2\theta_{13} \sin\theta_{23} \frac{\sin[(1-x) \Delta]}{(1-x)} \label{eq:pme_t2}\,,
\end{eqnarray}
and
$\alpha = \frac{\Delta m_{21}^{2}}{\Delta m_{31}^{2}}$, $x = \frac{2EV}{\Delta m_{31}^{2}}$, 
$\Delta = \frac{\Delta m_{31}^{2} L}{4E}$, and $ V = \pm 2\sqrt{2} G_{F} N_{e} $ is the matter potential (positive for neutrinos and negative for antineutrinos).
The only other relevant probability, $ P_{\mu\mu} $, is mildly dependent on \dcp.

\subsection{Analysis with a 35 kt unmagnetized \liar\ FD}\label{MH-mass}

As is evident from Fig.~\ref{fig:mh-dcp_700_350_unmag}, 
mass hierarchy resolution benefits significantly from having an ND.
But, note that the results with or without an ND are similar for regions of \dcp\  where the sensitivity is worse ($\dcp \in [45^\circ, 135^\circ]$ for the normal hierarchy and $\dcp \in [-135^\circ, -45^\circ]$ for the inverted hierarchy).

\begin{figure*}[tb]
	\centering
	\includegraphics[width=1.02\textwidth]{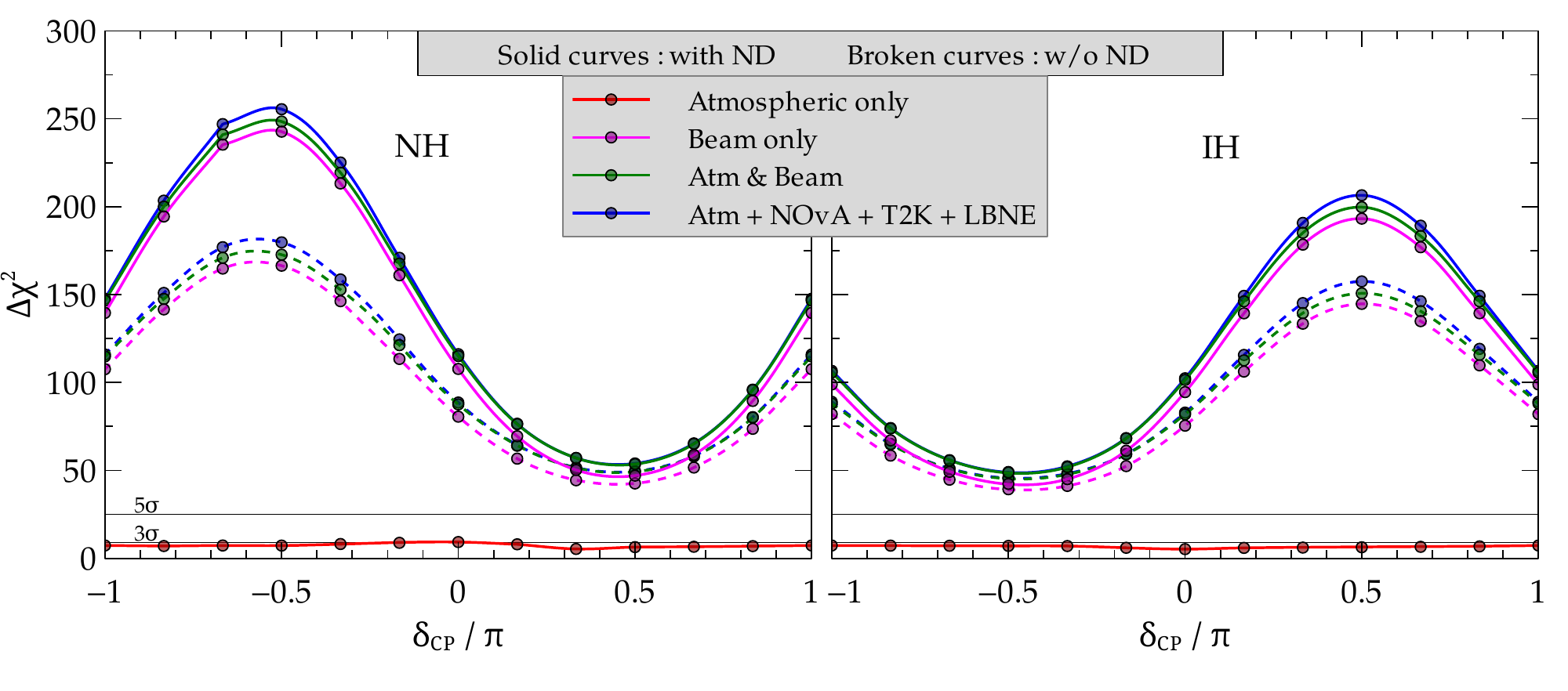}
	\caption{\label{fig:mh-dcp_700_350_unmag}Sensitivity to the mass hierarchy as a function of true $\delta_{\text{CP}}$ for a true normal hierarchy (NH) and a true inverted hierarchy (IH) with an 350~kt-yr exposure at the unmagnetized far detector configured with and without a near detector (ND). A run-time of 5 years each ($3\times 10^{21}$ protons on target) with a $\nu$ and $\anti{\nu}$ beam is assumed. The combined sensitivity with NO$\nu$A (15 kt TASD, 3 yrs. $\nu$ + 3 yrs. $\bar{\nu}$) and T2K (22.5 kt water cerenkov, 5 yrs. $\nu$) data is also shown.}
\end{figure*}

\begin{figure*}[bt]
	\centering
	%\begin{subfigure}{0.4\textwidth}
		%\centering
		\includegraphics[width=1.03\textwidth]{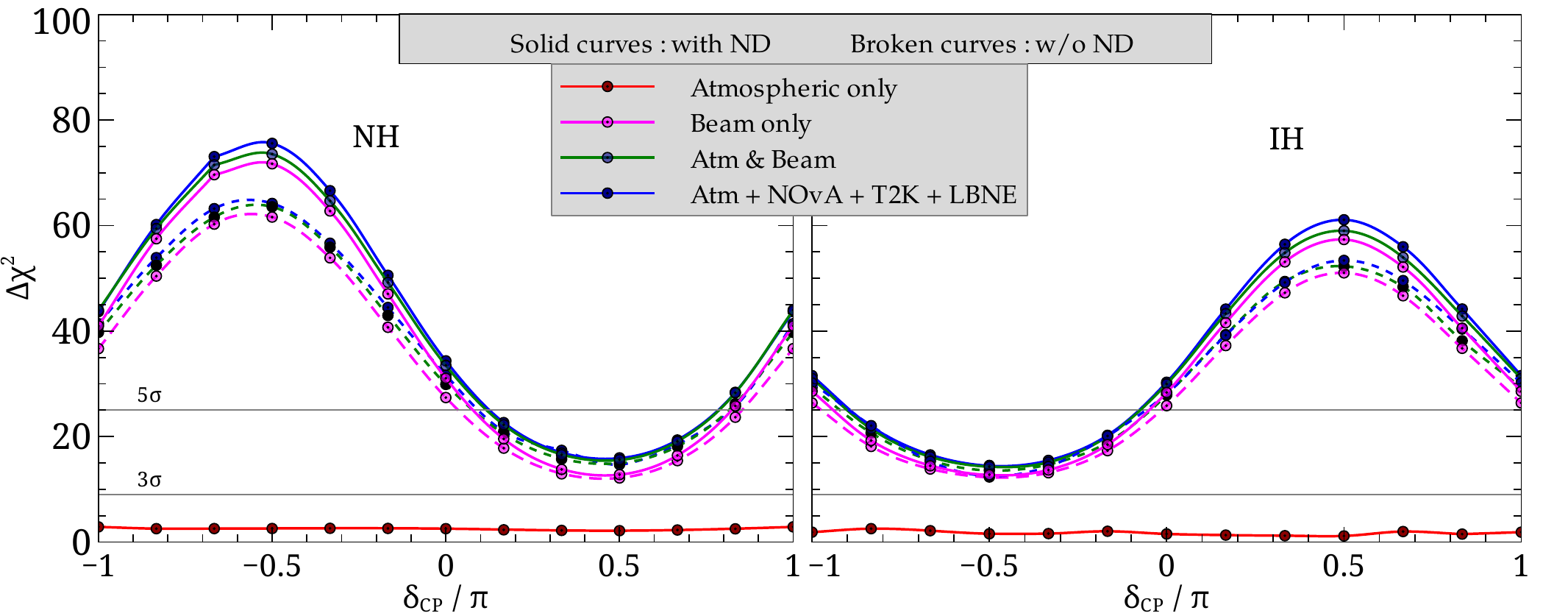}
	%\end{subfigure}
	\caption{\label{fig:mh-dcp_700_100_unmag}Similar to Fig.~\ref{fig:mh-dcp_700_350_unmag} but for a 100 kt-yr unmagnetized \liar\ FD.}
\end{figure*}

Since the wrong hierarchy can be excluded by the DUNE beam-only experiment to significantly more than $5\sigma$ with an unmagnetized \liar\ FD and an exposure of 350 kt-yrs. without the help of ND, the added contributions of both the atmospheric neutrinos, and the better signal and background systematics provided by an ND, are not essential for this measurement.

Similar conclusions vis-a-vis the near detector can be drawn for a 10 kt FD from Fig.~\ref{fig:mh-dcp_700_100_unmag}. For a 100~kt-yr exposure, the combined analysis resolves the hierarchy to more than $5\sigma$ for a large \dcp\ fraction, and to more than $3\sigma$ for all values of \dcp.

\subsection{Exposure analysis}
\label{sec:MH-exp}
\begin{figure*}[tb]
	\centering
	\includegraphics[width=1.01\textwidth]{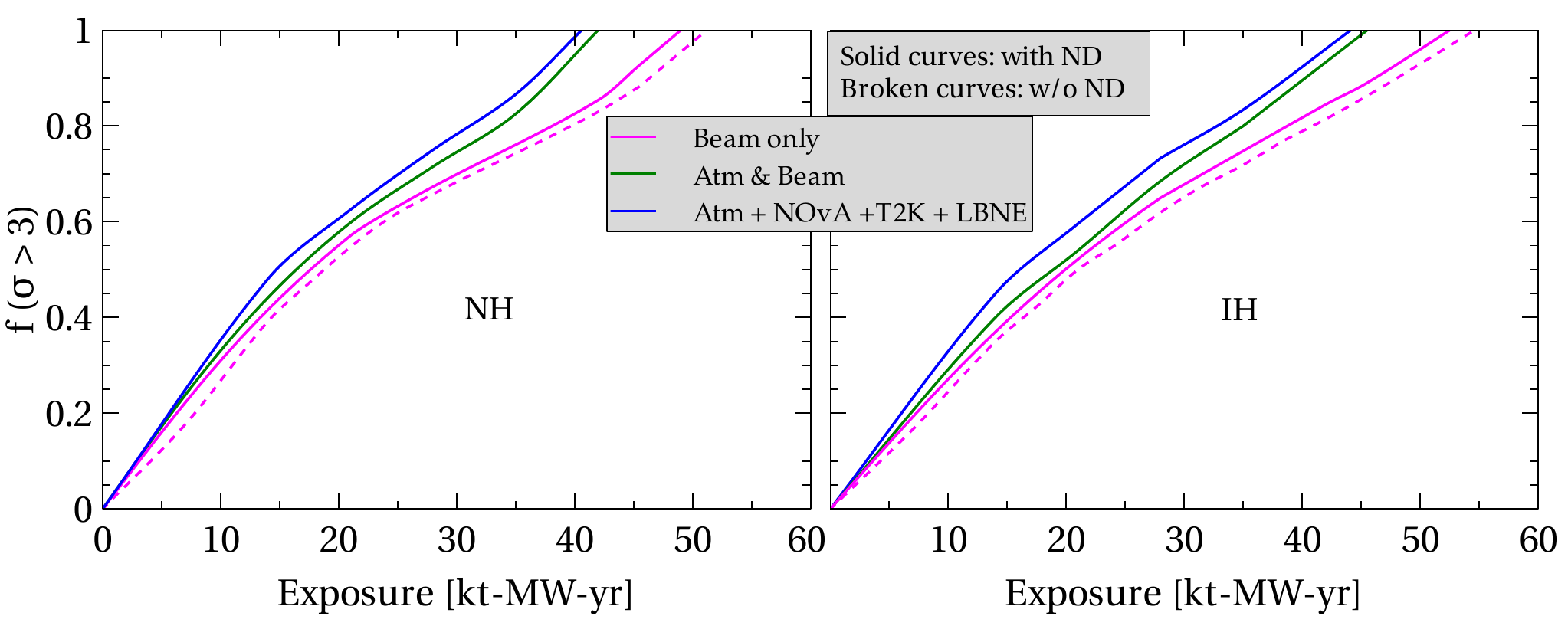}
	\caption{\label{fig:3sigma_mh}The fraction of CP phases for which  the sensitivity to the mass hierarchy exceeds $3\sigma$ as a function of DUNE exposure, for different unmagnetized detector configurations. The time exposure refers to calendar years for DUNE with $1.65\times 10^7$ seconds of uptime per year. The entire NO$\nu$A and T2K datasets are assumed to be available when DUNE starts taking data (and do not contribute to the exposure shown). }
\end{figure*}

We now evaluate the exposure needed to resolve the mass hierarchy for the entire range of \dcp.
In Fig.~\ref{fig:3sigma_mh}, we show the CP fraction (f($\sigma > 3$)) for which the sensitivity to 
mass hierarchy is greater than $3\sigma$, as a function of exposure. 
Salient points evident from Fig.~\ref{fig:3sigma_mh} are:

\begin{itemize}

\item For a beam only analysis, we see that a $3\sigma$ determination of the hierarchy for any \dcp\ value is possible with a roughly 50 kt-MW-yr exposure. This means the hierarchy can be resolved by a 35 kt FD and a 700~kW beam in  two years.

\item  A near detector does not reduce the exposure needed for a $3\sigma$ measurement. 

\item Information from atmospheric neutrinos reduces the exposure required to about 45~kt-MW-yr. 

\item A further combination with NO$\nu$A and T2K data provides minor improvement.

\end{itemize}

\subsection{Variation of systematics}
\label{sec:MH-exp-syst}
\begin{figure*}[tb]
	\centering
	\includegraphics[width=1.0\textwidth]{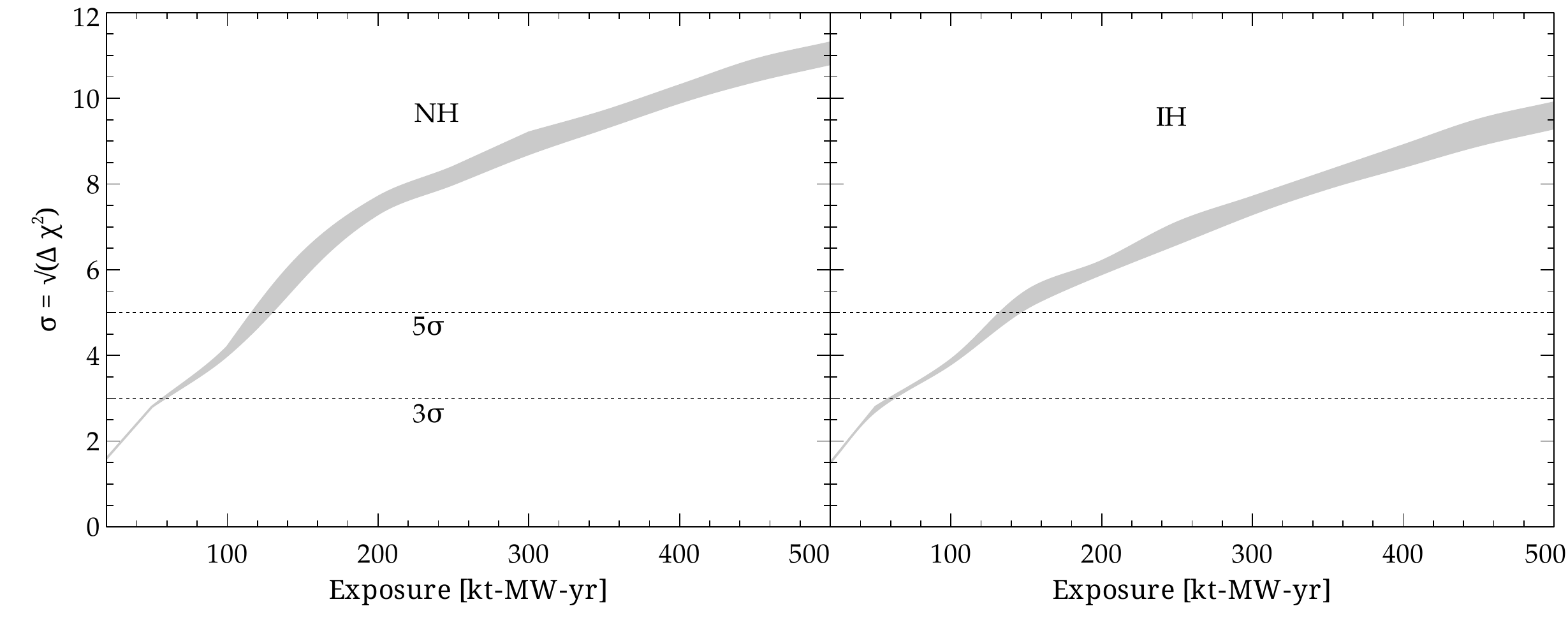}
	\caption{\label{fig:sigma_vs_exp_mh} Maximum sensitivity to the mass hierarchy for all values of $\dcp$ allowing  for different systematics 
	(see Sec.~\ref{sec:MH-exp-syst}), as a function of exposure. Only beam data (with both an FD and ND) have been considered.}
\end{figure*}
In Fig.~\ref{fig:sigma_vs_exp_mh}, we show the maximum sensitivity to the mass hierarchy
 for the entire $\dcp$ (true) space as a function of the 
exposure. We have allowed for variations in systematics for DUNE with an ND that are 3 times as large or small as those
in Table \ref{tab:syst}).
The width of the band produced by this procedure may  be considered
as a measure of the effect of systematics on the hierarchy sensitivity when an ND is used
(which seems to be the likely scenario in practice) along with an FD. We observe the following
features from Fig.\ \ref{fig:sigma_vs_exp_mh}:\\
\begin{itemize}
 \item For the NH (left panel), both $3\sigma$ and $5\sigma$ levels of sensitivity can be reached with an exposure of about 50 and 
 120 kt-MW-yr, respectively. This is consistent with Fig.~\ref{fig:3sigma_mh} wherein the solid magenta curve  in the left panel reaches unity at roughly 50~kt-MW-yr.
 
 \item For exposures below $\sim$ 20 kt-MW-yr, the sensitivity is not statistically significant ($\lesssim 1.5\sigma$).
 
 \item The variation of systematics has a small effect on the sensitivities for lower exposures ($\lesssim 100$ kt-MW-yr) and the effect gets slightly magnified for larger exposures, as evident from the widening of the bands.
 
 \item The hierarchy sensitivity for a true IH scenario (right panel of Fig.~\ref{fig:sigma_vs_exp_mh}) shows qualitatively similar behaviour as that for NH. 
\end{itemize}

% From Fig.\ \ref{fig:3sigma_mh}, discussed in sec.\ \ref{sec:MH-exp}, it is clear that 
% given a specific exposure, mass hierarchy can be resolved upto a maximum sensitivity for the entire $\dcp$ space.  

\subsection{Effect of magnetization}
\label{MH-mag}
\begin{figure*}[bt]
	\centering
	\includegraphics[width=1.01\textwidth]{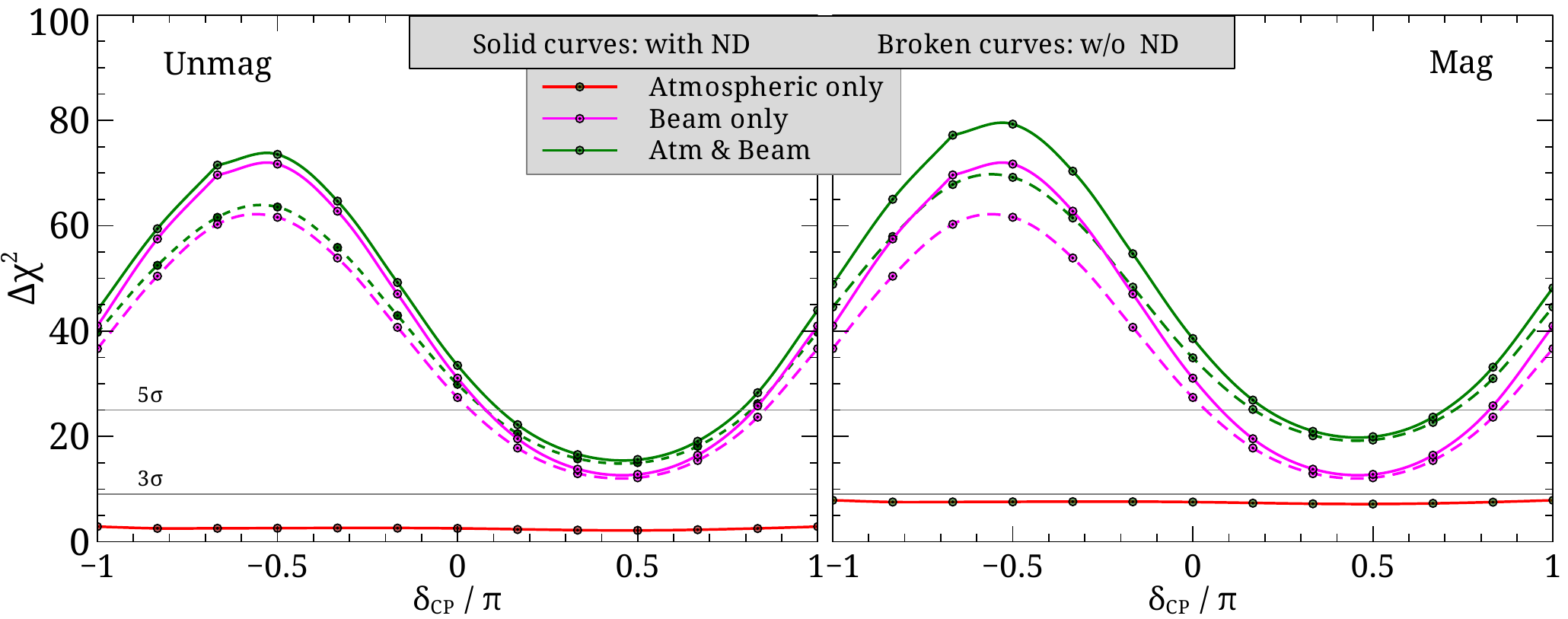}
	\caption{\label{fig:mh_nh_dlcp_mag_unmag}Sensitivity to the mass hierarchy with a 100 kt-yr exposure with a magnetized (mag) and unmagnetized (unmag) FD. The true hierarchy is normal.  }
\end{figure*}

In Fig.~\ref{fig:mh_nh_dlcp_mag_unmag}, we compare the sensitivity to the mass hierarchy of an unmagnetized and magnetized 100 kt-yr \liar\ FD for a true NH.
As discussed earlier, magnetizing the detector volume holds significance for the atmospheric neutrino analysis, since it allows the discrimination of neutrinos and antineutrinos in the flux.
Consequently, for the magnetized detector the atmospheric neutrinos alone contribute an almost $3\sigma$ sensitivity, thus also enhancing the combined sensitivity; the beam-only results remain unaffected by magnetization. 

\subsection{Some remarks on the results}

Some understanding of the qualitative nature of the results may be gleaned from considering the relevant expressions at the level of oscillation probabilities.

Ignoring systematic uncertainties, a {\it theoretical} event-rate dataset ($N^{\text{th}}$) and an {\it experimental} one ($N^{\text{exp}}$) can be used to define
\be \label{eq:chisq_analysis0}
\chisq \sim \sum_{i}\frac{(N^{\text{th}}(i) - N^{\text{exp}}(i))^{2}}{N^{\text{exp}}(i)}\,.
\ee
Since the event rate at energy $E$ in the $ e^- $ appearance channel is given by $ \pmue(E) \times \Phi(E) \times \sigma_\text{CC}(E)$, the sensitivity to the mass hierarchy follows the behavior of
\begin{equation}\label{eqn:pidefn}
	\pimh(\text{A}, \text{B}) = \left(\pmue^\text{A} - \pmue^\text{B}\right)^2,
\end{equation}
where A represents the assumed true hierarchy and B represents the test hierarchy.
The opposing natures of the sensitivities seen for the NH as true and IH as true scenarios respectively
can be related to the \dcp-dependent phase difference between $ \pimh(\text{NH}, \text{IH}) $ and
$ \pimh(\text{IH}, \text{NH}) $ at energies where the flux is high, \ie, 1.5--3.5 GeV; see Fig.~\ref{fig:fluxsec} in Appendix~B.
The \emph{oscillatory} nature of \dchisq\ with respect to \dcp\ in each case can be
traced back to Eq.\ \eqref{eqn:pidefn} too.

Because of the strong parameter space degeneracies involved in the appearance channel probability [Eq.\ \eqref{eq:pme}], neither the T2K nor the \nova\ experiments are capable of significantly improving the mass hierarchy sensitivities in their respective configurations.
It is apparent that the mass hierarchy study benefits immensely from the longer baseline as well as improved systematics of the DUNE set-up as compared to T2K and \nova.

Since the mass hierarchy will be determined at $3\sigma$ with relatively little
exposure (see Fig.\ \ref{fig:3sigma_mh}), henceforth, we assume the mass hierarchy to be known.
It is well known that studies of the octant degeneracy and CP violation benefit significantly from knowledge of the mass hierarchy.

\section{Octant degeneracy}
\label{sec:od}

We test the sensitivity to the $ \theta_{23} $ octant by using the true value to be equivalent to the present best-fits $\sin^{2}(\theta_{23}^{\text{true}}) = 0.427$ (0.613) for the lower (higher) octants (LO/HO), except for Fig.\ \ref{fig:th23-nnd}, where we show the sensitivities for a range of $ \theta_{23} $ values in both octants.
The \dchisq\ in each case represents the sensitivity to disfavoring the opposite octant when a particular choice of $ \theta^\text{true}_{23} $ is made.

The results for the octant analysis have one important feature --- the inverted hierarchy
scenario shows almost no variation in sensitivity with \dcp.
This follows from the well-known result that when neutrino masses are arranged
in an inverted hierarchy, the contribution to the \dchisq\ comes almost equally from
antineutrinos and neutrinos, but with nearly opposite values of \dcp, while if they conform to the
normal hierarchy, the neutrinos dominate over the antineutrinos and the overall result
largely traces the features of the neutrino-only \dchisq.

\subsection{Analysis with a 35 kt unmagnetized \liar~FD}
\label{sec:od-mass}
\begin{figure*}[tb]
	\centering
	\includegraphics[width=1.037\textwidth]{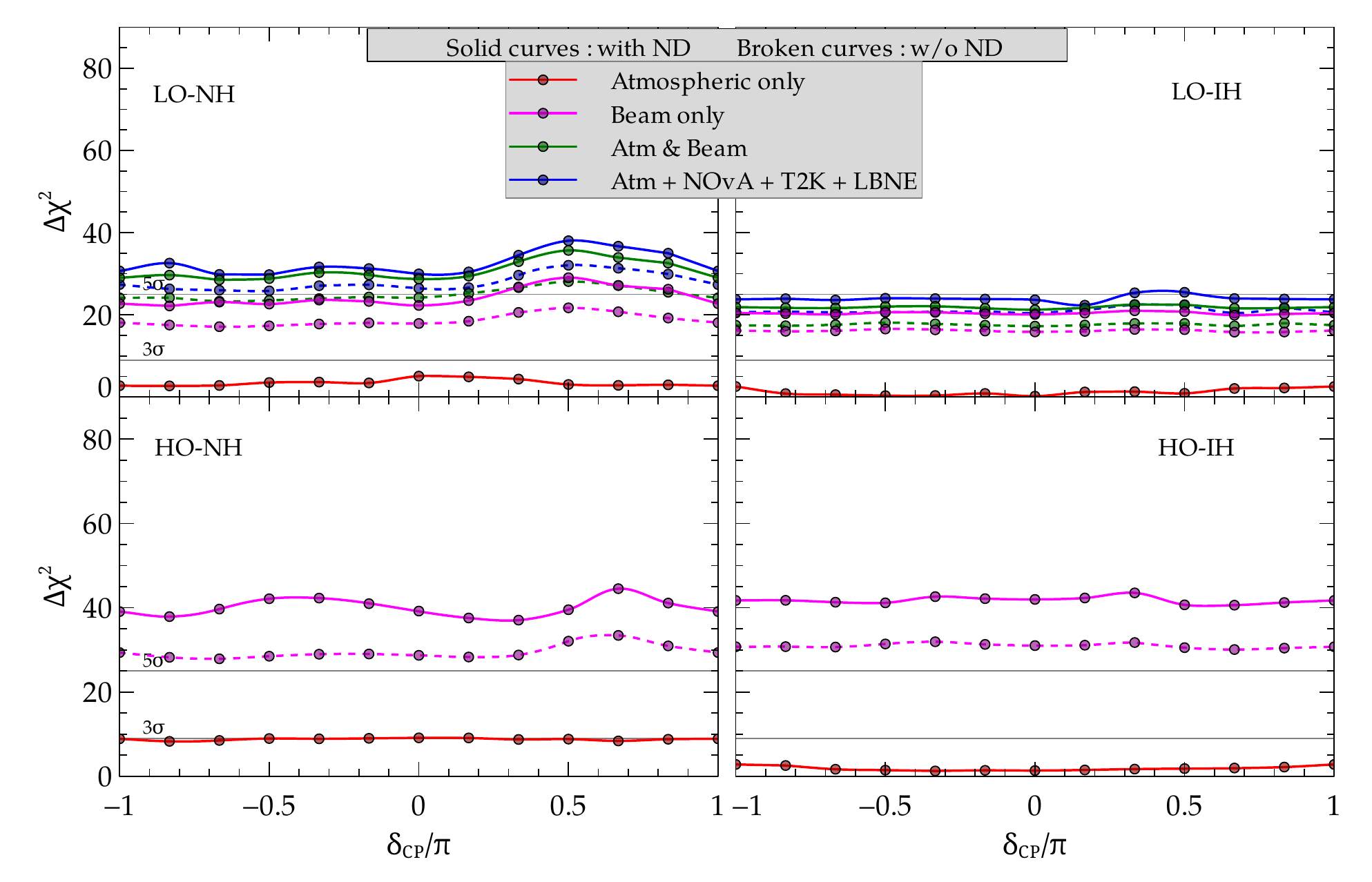}
	\caption{\label{fig:oct-dcp-th13-prior-nnd}Sensitivity to the octant of $\theta_{23}$ with
$\sigma(\sin^2 2\theta_{13}) = 0.05\times\sin^{2}2\theta_{13}$ and $\sin^{2}\theta_{23} = 0.427$ in the lower octant (LO) and $\sin^{2}\theta_{23} = 0.613$ in the higher octant (HO), for both hierarchies and a 350 kt-yr unmagnetized FD exposure configured with and without an ND. Representative results of the combined sensitivity with atmospheric data, and with NO$\nu$A and T2K are shown for the lower octant.
}
\end{figure*}

Figure~\ref{fig:oct-dcp-th13-prior-nnd} shows the sensitivity to the octant of $\theta_{23}$ for a given mass hierarchy.\footnote{The $\chi^{2}$ analysis converges to the minimum extremely slowly in the HO case of the combined setup. Hence, for results in this section, we only show representative plots for the combined setup for the LO case. We expect qualitatively similar results for the HO case.}
By and large, a beam only analysis with an ND provides better sensitivity than the combined analysis with atmospheric data without an ND. Only for the lower octant and normal hierarchy (LO-NH), do the sensitivities almost coincide. 

Figure~\ref{fig:th23-nnd} shows the octant sensitivity as a function of true $\theta_{23}$. No sensitivity is expected for $\theta_{23} = 45^{\circ}$.
 The sensitivity with an ND is slightly greater than the combined analysis with atmospheric data without an ND if the true hierarchy is inverted, and the converse is true for a normal hierarchy.

\begin{figure*}[tb]
	\centering
	\includegraphics[width=1.02\textwidth]{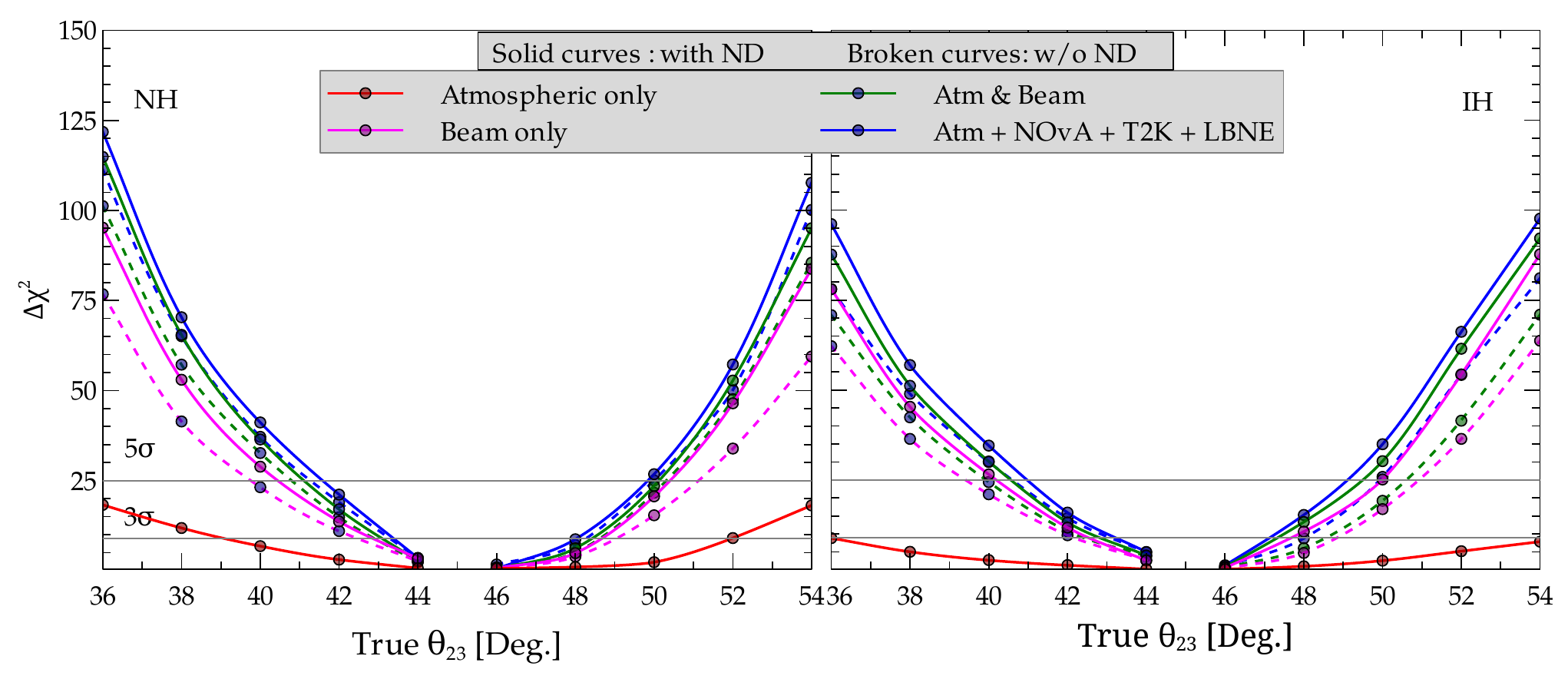}
	\caption{\label{fig:th23-nnd}Sensitivity to the octant for a 350~kt-yr unmagnetized FD exposure as a function of $\theta_{23}$ and with $\sigma(\sin^{2}2\theta_{13}) = 0.05\times \sin^{2}2\theta_{13}$. 
}
\end{figure*}

\subsection{Effect of $\theta_{13}$ precision}
\label{sec:od-prior}

\begin{figure*}[tb]
\centering
	\includegraphics[width=1.02\textwidth]{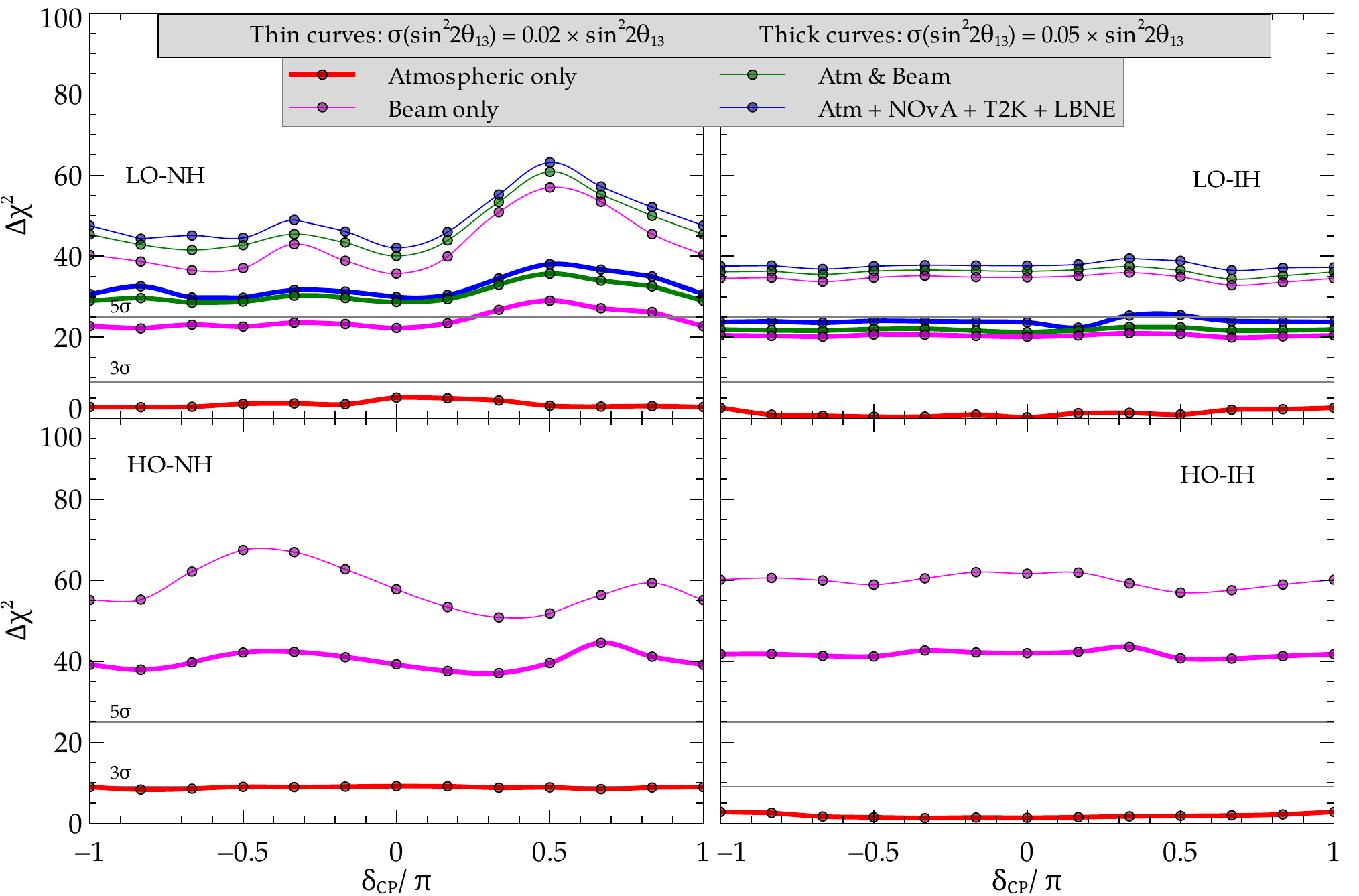}
	\caption{\label{fig:oct-dcp-th13-prior-350}Octant sensitivity for $\sigma(\sin^{2}2\theta_{13}) = 0.02\times \sin^{2}2\theta_{13}$ and $\sigma(\sin^{2}2\theta_{13}) = 0.05\times \sin^{2}2\theta_{13}$ 
	for a 350 kt-yr unmagnetized FD and an ND. Representative results of the combined sensitivity with atmospheric data, and with NO$\nu$A and T2K are shown for the lower octant.\protect \footnotemark[1]
}
\end{figure*}

Resolving the octant degeneracy depends greatly on the precision with which $\theta_{13}$ is known.
As Fig.~\ref{fig:oct-dcp-th13-prior-350} shows, a 2\% uncertainty on $\sin^2 2\theta_{13}$ significantly improves the octant sensitivity compared to a  5\% uncertainty on $\sin^{2}2\theta_{13}$.

\subsection{Exposure analysis}
\label{od-exposure}
\begin{figure*}[tb]
	\centering
	\includegraphics[width=1.02\textwidth]{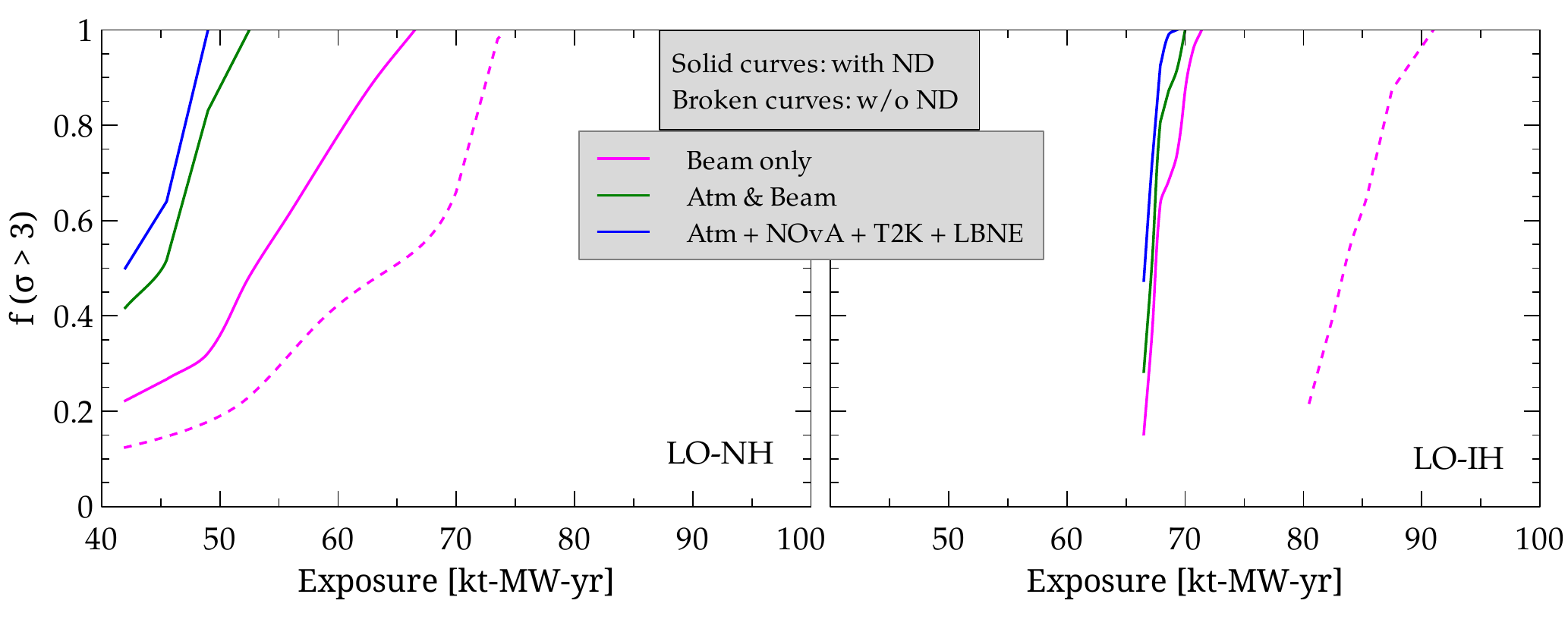}
	\caption{\label{fig:lo-exp}
The fraction of CP phases for which  the sensitivity to the octant exceeds $3\sigma$ as a function of exposure, for $\theta_{23}$ in the lower octant (LO)
and $\sigma(\sin^{2}2\theta_{13}) = 0.05\times\sin^{2}2\theta_{13}$.
}
\end{figure*}

Results of our exposure analysis are shown separately for the two octants
in Figs.~\ref{fig:lo-exp} and~\ref{fig:ho-exp}. We note the following:

\begin{itemize}

\item In the case of LO-NH, for exposures below 50~kt-MW-yr, f($\sigma > 3$) rises slowly with exposure for a beam-only experiment with or without a near detector; see Fig.~\ref{fig:lo-exp}. 
Above this exposure, the curves steepen and eventually the degeneracy is broken for all \dcp\ values for a 75~kt-MW-yr exposure. Thus, the octant will
be known at $3\sigma$ in one year after
the mass hierarchy is determined with a 35~kt detector and 700~kW beam.

\item The right panels of Figs.~\ref{fig:lo-exp} and~\ref{fig:ho-exp} show the IH case. The steepness of the curves can be understood from Fig.~\ref{fig:oct-dcp-th13-prior-nnd} which shows that $\Delta$\chisq\ is almost independent of \dcp\ for the IH.
As the exposure is increased, $\Delta$\chisq\ increases, and above a critical
exposure, the octant degeneracy is broken at $3\sigma$ for almost all values of \dcp\ with a small increment in exposure.

\item From Fig.~\ref{fig:ho-exp}, we observe that
the lower octant  can be ruled out at $3 \sigma$ for the entire
 \dcp\ parameter space with less than a 40~kt-MW-yr exposure, with or without a near detector.

\end{itemize}

\begin{figure*}[t]
	\centering
	\includegraphics[width=1.025\textwidth]{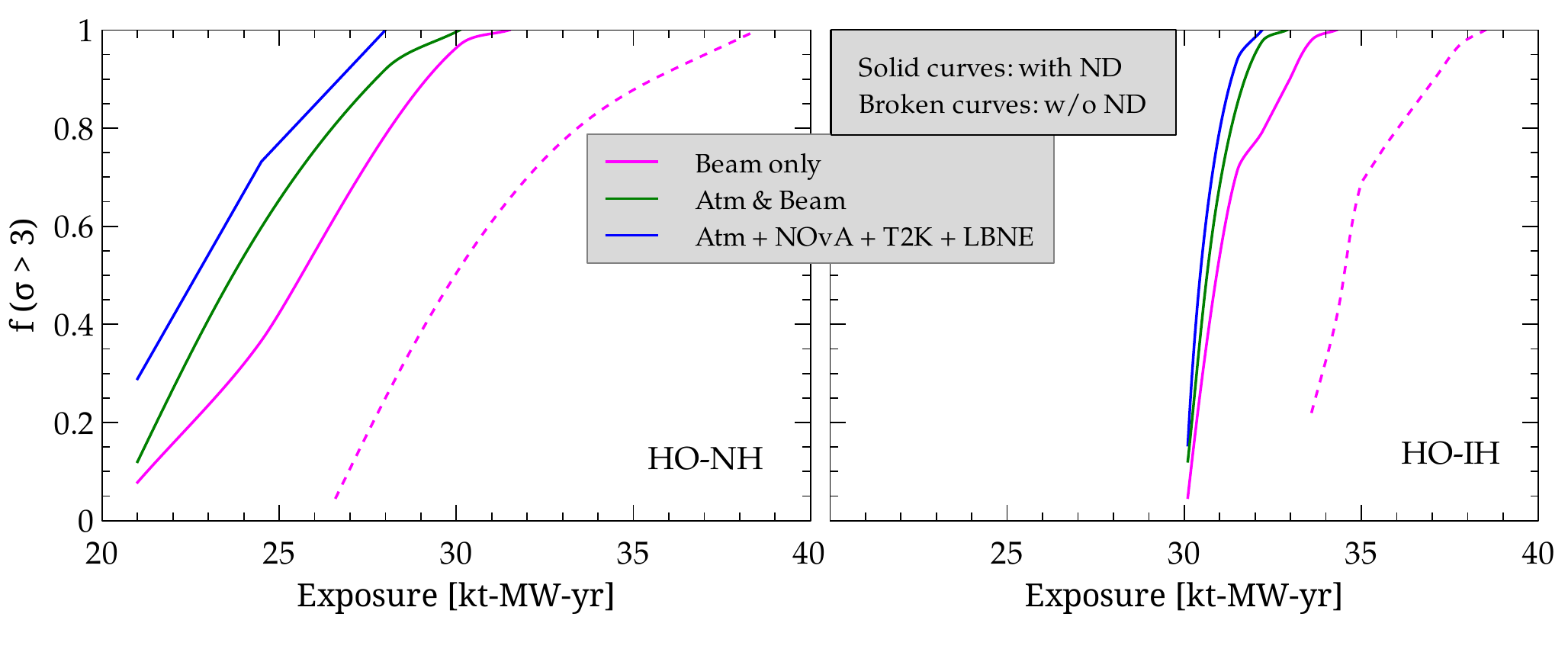}
	\caption{\label{fig:ho-exp}
The fraction of CP phases for which  the sensitivity to the octant exceeds $3\sigma$ as a function of exposure, for $\theta_{23}$ in the higher octant (HO)
and $\sigma(\sin^{2}2\theta_{13}) = 0.05\times\sin^{2}2\theta_{13}$.}
\end{figure*}

\subsection{Variation of systematics}
\label{sec:octant-exp-syst}
\begin{figure*}[tb]
	\centering
	\includegraphics[width=1.02\textwidth]{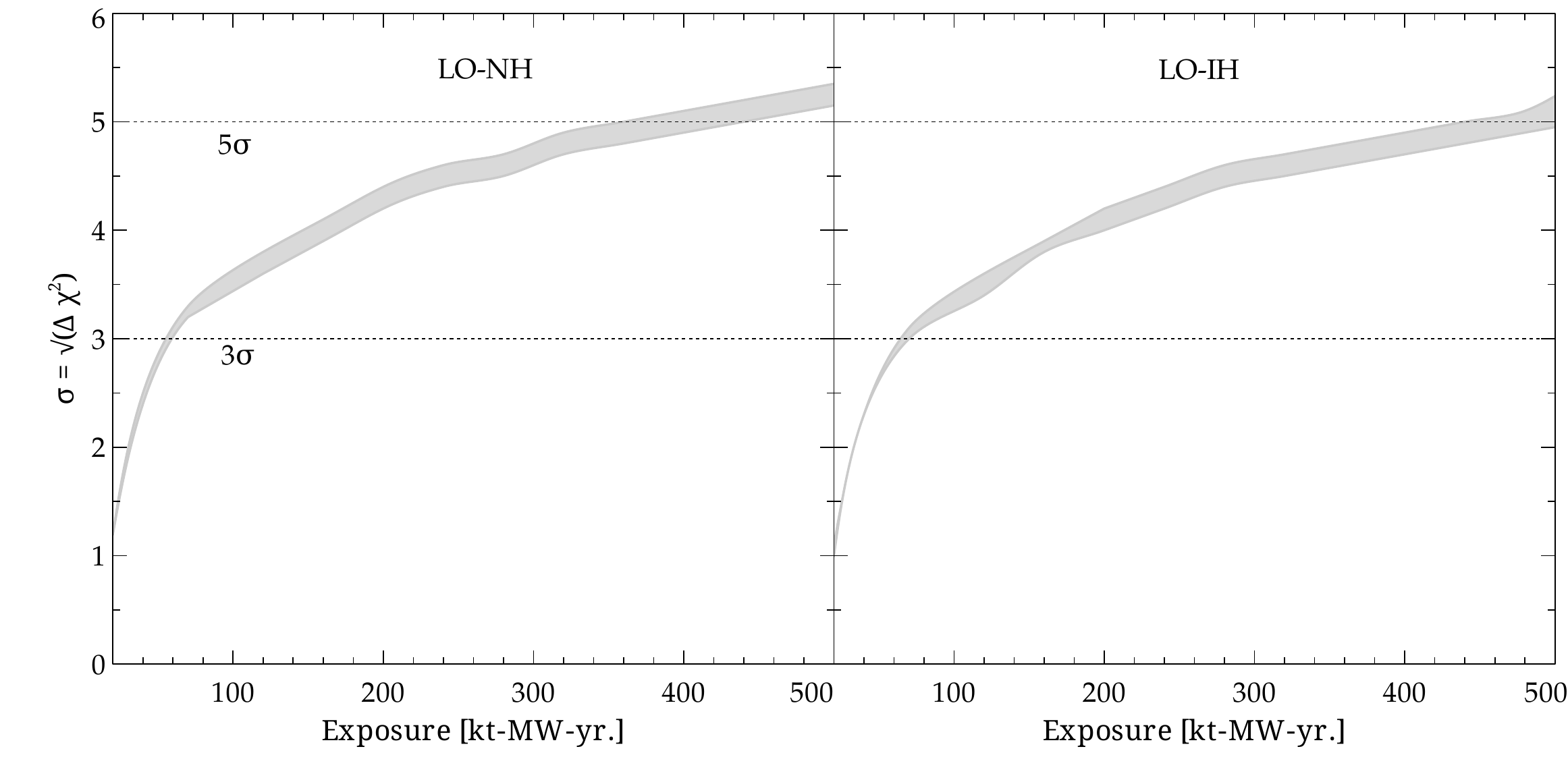}
	\caption{\label{fig:sigma_vs_exp_octant}Similar to Fig.~\ref{fig:sigma_vs_exp_mh}, but for the maximum sensitivity to a true lower octant (LO).}
\end{figure*}

In Fig.~\ref{fig:sigma_vs_exp_octant}, we plot the maximum sensitivity to the octant that can be achieved for all values of $\dcp$. We express the sensitivity as a band, obtained by varying the systematics as described in Sec.~\ref{sec:MH-exp-syst}). We note from Fig.~\ref{fig:sigma_vs_exp_octant} that,\\
\begin{itemize}
 \item The $3\sigma$ sensitivity level can be achieved with $\sim 70$~kt-MW-yr for both hierarchies. This is consistent with Fig.\ \ref{fig:lo-exp}. 
 
 \item For the NH, it takes less ($\sim 400$~kt-MW-yr) exposure to obtain $5\sigma$ sensitivity compared to the true IH scenario ($\sim 500$~kt-MW-yr).
 
 \item The variation of systematics has a negligible effect on the sensitivity for exposures $\lesssim 70$ kt-MW-yr until $3\sigma$ sensitivity is reached. Thereafter, to achieve the $5\sigma$ level, the sensitivities get slightly more affected by systematics.  
\end{itemize}

\subsection{Effect of magnetization}
\label{od-mag}

\begin{figure*}[t]
	\centering
	\includegraphics[scale=0.5]{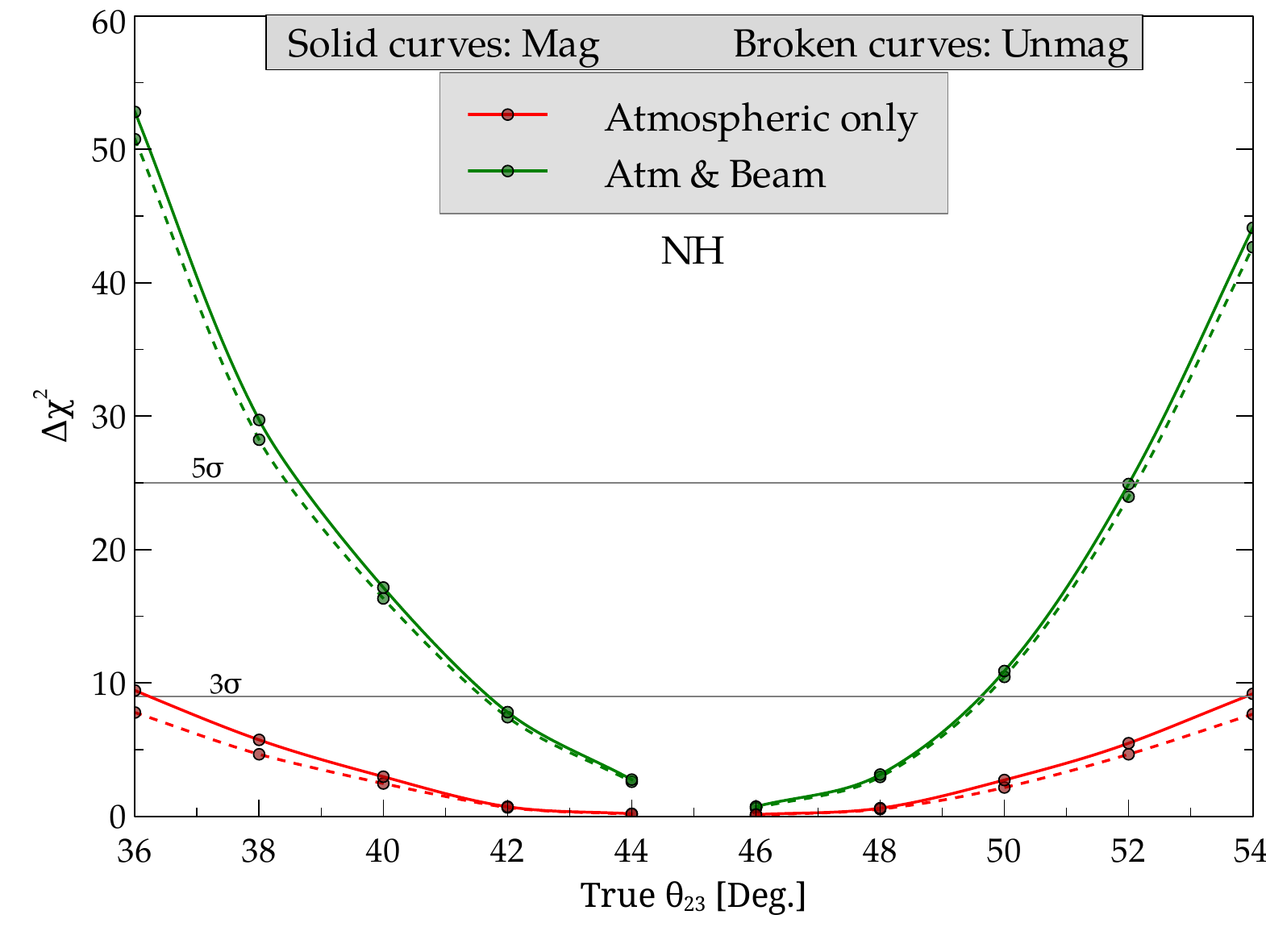}
	\caption{\label{fig:th23-nomag}Octant sensitivity of a magnetized and an unmagnetized detector with a 100~kt-yr exposure.
 The true hierarchy is assumed to be normal.}
\end{figure*}

We see from Fig.~\ref{fig:th23-nomag}, that magnetizing the detector has almost no effect on the sensitivity to the octant. From a practical standpoint, it is unlikely that a 35~kt FD would be magnetized, with no consequence for octant sensitivity.

\paragraph*{}
It is apparent that the octant resolution benefits significantly from the presence of a calibrating ND.
As Figs.\ \ref{fig:lo-exp} and \ref{fig:ho-exp} show, the consequent improvement in statistics reduces the runtime required for the achievement of a $ 3\sigma $ significant resolution by at least 8 yrs (for NH, both octants) and as much as 20 yrs for the LO-IH scenario.

In contrast,  the benefit of adding the atmospheric flux is modest, except in the case of LO-NH where a combination of the atmospheric and beam sensitivities can drive the resolution to almost $ 5\sigma $ significance despite the absence of the ND.
While the atmospheric contribution is also large in the case of HO-NH, in this case the beam-only epxeriment is already capable of resolving the degeneracy to more than a $ 3\sigma $ level by itself even without the ND.
In the latter case, therefore, the additional expenditure that would be inevitably involved in building the FD underground, would not be justifiable.

The octant degeneracy resolution also greatly benefit from investment in increasing the FD volume to 35 kt from 10 kt, as is evident from the Figs.\ \ref{fig:lo-exp} and \ref{fig:ho-exp}.

\section{CP violation}
\label{sec:cp}

Of the six oscillation parameters, \dcp\ is the least well known.
Part of the reason for this was the difficulty in experimentally determining the value of $ \theta_{13} $.
With reactor experiments over the last three years having made significant progress toward the precision determination of the latter, and it being established by now that the value of $ \theta_{13} $ is non-zero by a fair amount, the precision determination of \dcp\ in a future experiment should be possible.

In the following we study the sensitivity of the DUNE to CP-violation in the neutrino sector brought about by a non-zero \dcp\ phase.
To determine the \dchisq\ that represents the experiment's sensitivity to CP-violaion, we assume a test \dcp\ value of 0 (or $\pi$) and compute the \dchisq\ for any non-zero (or $\neq \pi$) true \dcp.
Since the disappearance channel probability \pmumu\ is only mildly sensitive to the \dcp, CP-violation in the neutrino sector can only be studied by experiments sensitive to the appearance channel $ \nu_\mu \to \nu_e $.
It is obvious, given the nature of the latter channel's probability \pmue, that the maximum sensitivity will be due to values of true \dcp\ close to odd multiples of $ \pi/2 $. 

Due to the non-zero value of $ \theta_{13}$ being now established, other experiments sensitive to the $ \nu_\mu \to \nu_e $ appearance channel, including the T2K and \nova, are also strongly poised to look for CP-violation.
Consequently, this is one study where combining data from DUNE, T2K and \nova\ proves to be significantly beneficial.

\subsection{Analysis with a 35 kt unmagnetized \liar~FD}
\label{cp-mass}

\begin{figure*}[tb]
	\centering
	%\begin{subfigure}%{0.49\textwidth}
		%\centering
		\includegraphics[width=1.02\textwidth]{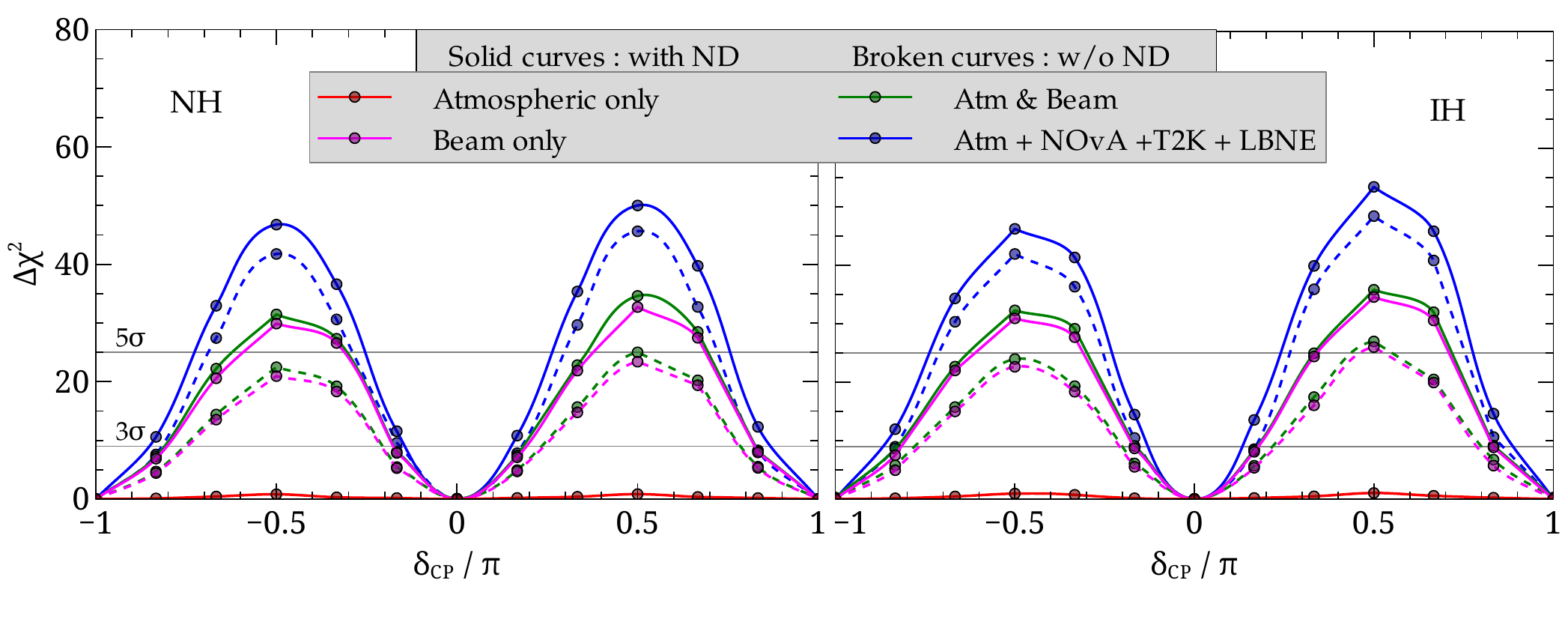}
	%\end{subfigure}
	%\begin{subfigure}%{0.49\textwidth}
		%\centering
		%\includegraphics[scale=0.45]{UG_ND_NND_IH_CPV_PRIOR_0_05}
	%\end{subfigure}
	\caption{\label{fig:cpv2}Sensitivity to CP violation for a 350 kt-yr  unmagnetized FD exposure
assuming $\sigma(\sin^{2}2\theta_{13}) = 0.05\times\sin^{2}2\theta_{13}$. }
\end{figure*}

To study CP violation, reduced systematics courtesy the placement of an ND proves to be beneficial (Fig.~\ref{fig:cpv2}).
Maximal CP violation can be ruled out at more than $5\sigma$ by a beam only analysis with
a 350 kt-yr exposure in conjunction with the ND.
However, $ 5\sigma $ resolution toward ruling out maximal CP-violation can even be achieved despite the absence of an ND by combining results from the T2K, \nova\ and the DUNE.

\subsection{Exposure analysis}
\label{sec: cp-exp}
\begin{figure*}[t]
	\centering
%	\begin{subfigure}{0.49\textwidth}
%		\centering
		\includegraphics[width=1.02\textwidth]{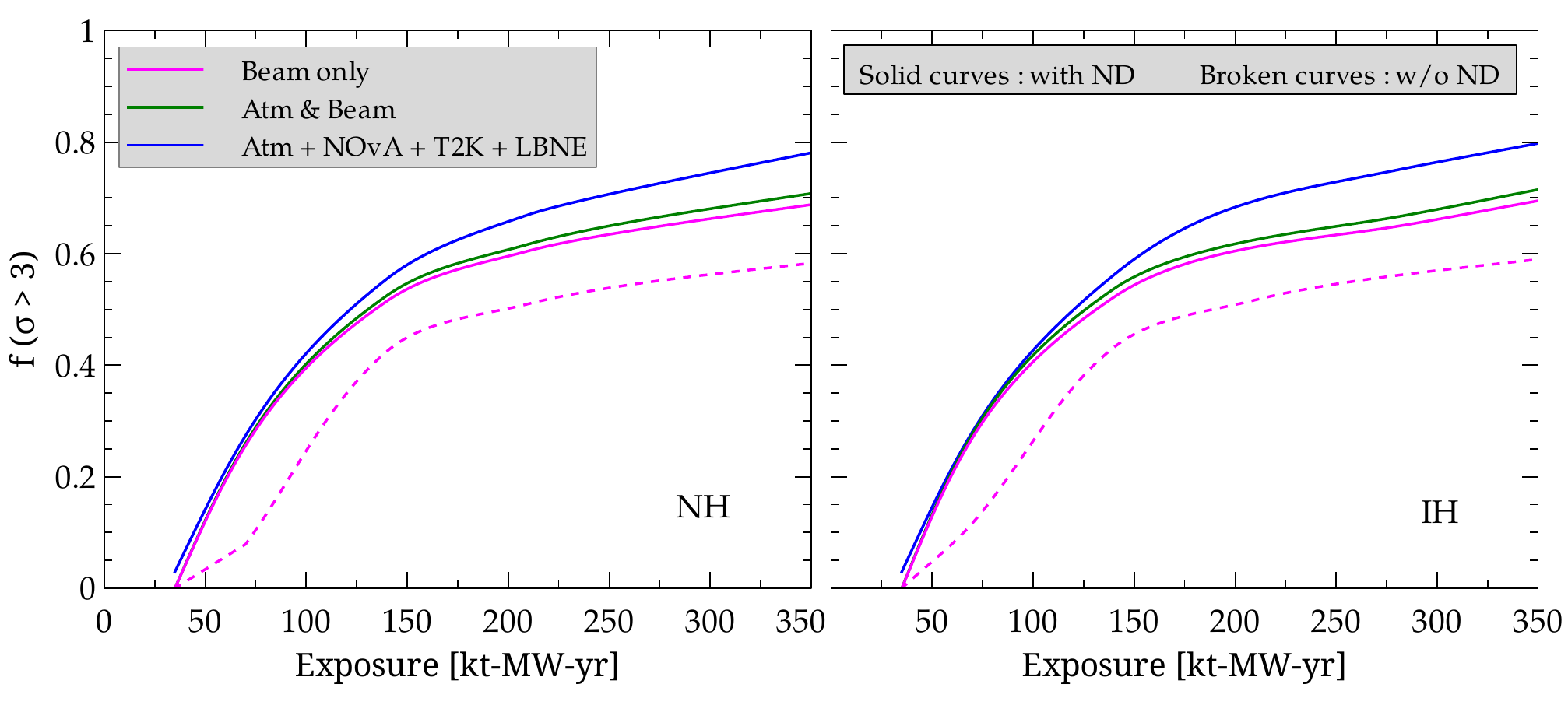}
%	\end{subfigure}
%	\begin{subfigure}{0.49\textwidth}
%		\centering
%		\includegraphics[scale=0.5]{ChiSq-Th23-IH-NND}
%	\end{subfigure}
	\caption{\label{fig:3sigma_cpv}
The fraction of CP phases for which  the sensitivity to CPV exceeds $3\sigma$ as a function of exposure.
}
\end{figure*}

In Fig.~\ref{fig:3sigma_cpv}, we show the CP fraction for which CP violation
can be established at $3\sigma$. Needless to say,
the CP fraction has to be less than unity since even an almost ideal experiment 
cannot exclude CP violating values of the phase that are close to the CP conserving values, $0$ and $\pi$. In the context of CP violation, the CP fraction
is a measure of how well an experiment can probe small CP violating effects.
From Fig.~\ref{fig:3sigma_cpv}, we find:

\begin{itemize}

\item There is no sensitivity to CP violation at the $3\sigma$ level for exposures smaller than about 35~kt-MW-yr. The sensitivity gradually
 increases with exposure and the CP fraction for which $3\sigma$ sensitivity is achieved approaches 0.4 (without an ND) and 0.5 (with an ND) for a 125 kt-MW-yr exposure. The CP fraction plateaus to a value below 0.8
for an exposure of 350~kt-MW-yr with all data combined.
 
\item A near detector certainly improves the sensitivity to CP violation.

\end{itemize}

\subsection{Variation of systematics}
\label{sec:cpv-exp-syst}
\begin{figure*}[tb]
	\centering
	\includegraphics[width=1.01\textwidth]{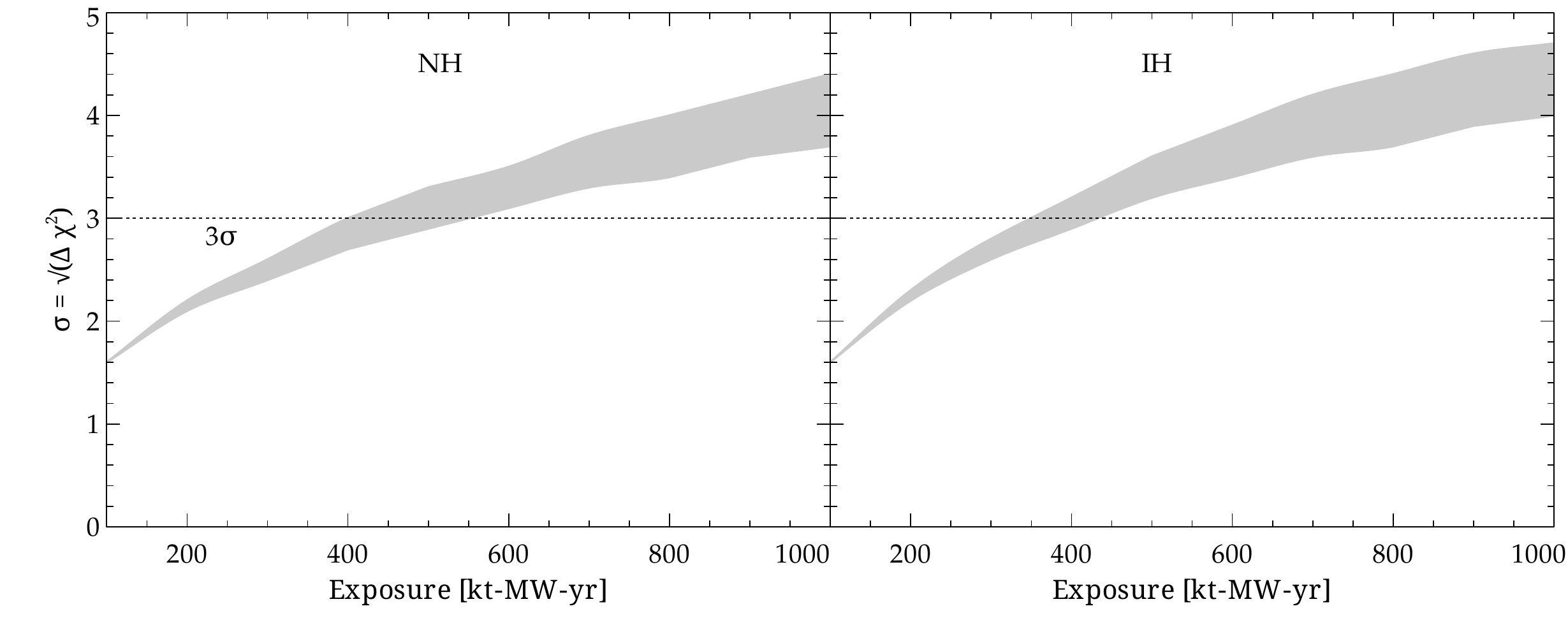}
	\caption{\label{fig:sigma_vs_exp_cpv}Similar to Figs.~\ref{fig:sigma_vs_exp_mh} and \ref{fig:sigma_vs_exp_octant}, but for the maximum sensitivity to CP violation for 70\% of the $\dcp$ parameter space.}
\end{figure*}
 Figure~\ref{fig:sigma_vs_exp_cpv} shows the maximum sensitivity to CP violation that can be achieved for $70\%$ of the $\dcp$ parameter space. As in Figs.~\ref{fig:sigma_vs_exp_mh} and~\ref{fig:sigma_vs_exp_octant}, we show the sensitivity as a band on varying the systematics. The notable features of Fig.~\ref{fig:sigma_vs_exp_cpv} are,\\
\begin{itemize}
  \item To resolve CP violation at the level of $3\sigma$ for $70\%$ region of the $\dcp$ space, a fairly long exposure is needed. For NH, it is roughly $400-500$~kt-MW-yr. while for IH it is $350-450$~kt-MW-yr. depending on the systematics.)

  \item For such long exposures, the sensitivity band becomes appreciably wide indicating a strong dependence on the
  systematics. In comparison, the sensitivities to the mass hierarchy and octant were less dependent on systematics since the corresponding exposures were smaller. This reinforces the need for an ND.
  
\end{itemize}

\subsection{Effect of magnetization}
\label{sec: cp-mag}
As can be seen from Fig.~\ref{fig:cpv2}, the sensitivity of atmospheric neutrinos to CPV is negligible, hence magnetizing the detector does not help.

\paragraph*{}
It is obvious that CP-violation is the study that stands to benefit most from the combination of results from the T2K, \nova\ and the DUNE.
Even potentially low sensitivity to maximal CP-violation due to the absence of ND can be overcome by the combination of \chisq\ data from the three epxeriments.
However, a large volume FD (35 kt) for the DUNE is almost certainly an absolute necessity, if any sensitivity to CP-violation has to be detected within a reasonable time frame, irrespective of the  benefits of combining results from other experiments such as the T2K and \nova.

The atmospheric neutrino flux has no role to play in the resolution of this physical problem.

\section{Summary}

We considered the Deep Underground Neutrino Experiment as either a 10~kt or 35~kt \liar\ detector situated underground at the Homestake mine and taking data in a high intensity neutrino beam for 5 years and in an antineutrino beam for another 5 years. 
For the 35~kt detector, we find that reduced systematic uncertainties afforded by a near detector greatly benefit the sensitivity to CP violation. However, a near detector provides only modest help with the octant degeneracy and is not necessary for the determination of the mass hierarchy since the sensitivity without a near detector is well above $5\sigma$. Since magnetization is not currently  feasible for a 35~kt detector,  we only considered this possibility for a 10~kt detector. While the sensitivity to the mass hierarchy from atmospheric neutrinos gets enhanced to almost $3\sigma$,
the  combined beam and atmospheric data is not much affected by magnetization. Also, magnetizing the detector does not help improve the sensitivity to the octant or to CP violation.

One thing is clear. A 35~kt DUNE will break all remaining vestiges of the eight-fold degeneracy that plagues long-baseline beam experiments~\cite{degen} and will answer all the questions it is designed to address.

\section*{Acknowledgments}
AB, AC, RG and MM  thank Pomita Ghoshal and Sanjib Mishra for discussions. This work was supported by 
US DOE grants DE-FG02-95ER40896 and DE-SC0010504. RG acknowledges the 
support of the  XI Plan Neutrino Project under DAE.

\section*{Appendix}
%\section{Numerical Procedure}
%label{sec:Num}
\section{Atmospheric neutrino analysis}
\label{sec:num-atm}
The simulation of atmospheric neutrino events and the subsequent $\chi^{2}$ analysis is carried out by means of a \verb-C++- program. Our method is described below.

\subsection{Event simulation}
\label{sec:num-atm-event}

The total number of CC events is obtained by  folding the relevant
incident neutrino fluxes with the appropriate disappearance and appearance probabilities, relevant CC cross sections, and the detector efficiency, resolution, mass, and
 exposure time. 
For our analysis, we consider neutrinos with energy in the range 1$-$10~GeV in 10 uniform bins,
and the cosine of the zenith angle $\theta$ in the range $-1.0$ to $-0.1$ in 18 bins. 
The $\mu^-$ event rate in an energy bin of width $\mathrm{dE}$ and in a solid angle bin of width ${\mathrm{d \Omega}}$ is,

\be \label{eq:muevent}
\rm{ \frac{d^2 N_{\mu}}{d \Omega \;dE} = \frac{1}{2\pi} \left[
\left(\frac{d^2 \Phi_\mu}{d \cos \theta \; dE}\right) P_{\mu\mu} +
\left(\frac{d^2 \Phi_e}{d \cos \theta \; dE}\right)
P_{e\mu}\right] \sigma_{CC} D_{eff} } \,.
\ee

Here ${\mathrm{\Phi_{\mu}}}$ and ${\mathrm{\Phi_{e}}}$ are the $\nu_\mu$ and ${\mathrm{\nu_e}}$ atmospheric fluxes, $P_{\mu\mu}$ and $P_{e\mu}$ are disappearance and appearance probabilities in obvious notation, $\rm{\sigma_{CC}}$ is the total CC cross section and $\rm{D_{eff}}$ is the detector efficiency. The $\mu^+$ event rate is similar to the above expression with the fluxes, probabilities and cross sections replaced by those for antimuons. Similarly, the ${\mathrm{e^-}}$ event rate in a specific energy and zenith angle bin is
\be \label{eq:eevent}
\mathrm{ \frac{d^2 N_e}{d \Omega \;dE} = \frac{1}{2\pi}
\left[\left( \frac{d^2 \Phi_{\mu}}{d \cos \theta \; dE}\right)
P_{\mu e} + \left(\frac{d^2 \Phi_e}{d \cos \theta \; dE}\right)
P_{e e} \right] \sigma_{CC} D_{eff} } \,,
\ee

with the ${\mathrm{e^+}}$ event rate being expressed in terms of
antineutrino fluxes, probabilities and cross sections. 

We take into account the smearing in both energy and
zenith angle, assuming a Gaussian form for the resolution
function, R. For energy, we use,
\be \label{eq:esmear}
\mathrm{ R_{E}(E_t,E_m) = \frac{1}{\sqrt{2\pi}\sigma} 
\exp\left[-\frac{(E_m - E_t)^2}{2 \sigma^2}\right]}\,.
\ee
Here, $\rm{E_m}$ and $\rm{E_t}$ denote the measured and true values
of energy respectively. The smearing width $\sigma$ is a function
of $\rm{E_t}$.

The smearing function for the zenith angle is a bit more complicated because the direction of the incident neutrino is specified by two variables: the polar angle ${\rm{\theta_t}}$ and the azimuthal angle ${\rm{\phi_t}}$. We denote both these angles together by ${\rm{\Omega_t}}$.
The measured direction of the neutrino, with polar angle ${\rm{\theta_m}}$ and azimuthal
angle ${\rm{\phi_m}}$, which together we denote by ${\rm{\Omega_m}}$, is expected to be within a cone of half angle $\Delta \theta$ of the true direction. The angular smearing is done in a small cone whose axis is given by the direction ${\rm{\theta_t, \phi_t}}$. The set of directions within the cone have different polar angles and azimuthal angles. Therefore, we need to construct a smearing function which takes into account the change in the azimuthal coordinates as well. Such an angular smearing function is given by,
\be \label{eq:anglesmear}
\rm{ R_{\theta}(\Omega_t, \Omega_m) = N \exp \left[ - \frac{(\theta_t -
\theta_m)^2 + \sin^2 \theta_t ~(\phi_t - \phi_m)^2}{2 (\Delta
\theta)^2} \right] } \,,
\ee
where N is a normalisation constant. 

Now, the \numu\ event rate with the smearing factors taken into account is given by,
\be \label{eq:eventrate}
\rm{ \frac{d^2 N_{\mu}}{d \Omega_m ~dE_m} = \frac{1}{2\pi}
\int \int dE_t~d\Omega_t~ R_{EN}(E_t,E_m)~
R_{\theta}(\Omega_t,\Omega_m)\left[\Phi_{\mu}^d \; P_{\mu\mu} +
\Phi_{e}^{d} \;P_{e\mu} \right]\sigma_{CC} D_{eff} }\,,
\ee
and similarly for the \nue\ event rate.
We have introduced the notation,
\[
\rm{(d^2 \Phi/d \cos \theta \;dE)_{\mu,e} \equiv \Phi_{\mu,e}^d}.
\]

Since $\rm{R_{EN}(E_{t}, E_{m})}$ and $\rm{R_{\theta}(\Omega_{t}, \Omega_{m})}$ are Gaussian, they can easily be integrated over the true angle $\Omega_{t}$ and the true energy $\text{E}_{\text{t}}$. Then,  integration over the measured energy $\text{E}_{\text{m}}$ and  measured angle $\Omega_{{m}}$ is carried out using the VEGAS Monte Carlo Algorithm.

\subsection{\chisq\ analysis}
\label{num-atm-chi}

The computation of \chisq\ for a fixed set of parameters is performed using the method of pulls. This method allows us to take into account the various statistical and systematic uncertainties in a straightforward way. The flux, cross sections and other systematic uncertainties are included by allowing these inputs to deviate from their standard values in the computation of the expected rate in the $\text{i-j}^{\text{th}}$ bin, ${\mathrm{ N^{th}_{ij} }}$. Let the ${\mathrm{
k^{th} }}$ input deviate from its standard value by ${\mathrm{
\sigma_k \;\xi_k }}$, where ${\mathrm{ \sigma_k }}$ is its
uncertainty. Then the value of ${\mathrm{ N^{th}_{ij} }}$ with the
modified inputs is
\begin{equation} \label{eqn:cij}
{\mathrm{ N^{th}_{ij} =  N^{th}_{ij}(std) + \sum^{npull}_{k=1}\;
c_{ij}^k \;\xi_k }} \,,
\end{equation}
where ${\mathrm{ N^{th}_{ij}(std) }}$ is the expected rate in the
$\text{i-j}^{\text{th}}$ bin calculated with the standard values of the
inputs and npull is the number of sources of uncertainty, which is 5 in
our case.
The ${\mathrm{ \xi_k }}$'s are called the \emph{pull} variables and
they determine the number of ${\mathrm{ \sigma}}$'s by which the
${\mathrm{ k^{th} }}$ input deviates from its standard value. In
Eq.~(\ref{eqn:cij}), ${\mathrm{ c_{ij}^k }}$ is the change in 
${\mathrm{N^{th}_{ij} }}$ when the ${\mathrm{ k^{th} }}$ input is changed by
${\mathrm{ \sigma_k }}$ (\ie~by 1 standard deviation). Since the
uncertainties in the inputs are not very large, we only consider changes in ${\mathrm{
N^{th}_{ij} }}$ that are linear in ${\mathrm{ \xi_k }}$. Thus we
have the modified $\chi^2$,
\begin{equation}  \label{eqn:chisq}
\mathrm{ {\chi^2(\xi_k)} = \sum_{i,j}\;
\frac{\left[~N_{ij}^{th}(std) \;+\; \sum^{npull}_{k=1}\;
c_{ij}^k\; \xi_k - N_{ij}^{ex}~\right]^2}{N_{ij}^{ex}} +
\sum^{npull}_{k=1}\; \xi_k^2  }\,,
\end{equation}
where the additional ${\rm{\xi_k^2}}$-dependent term is the penalty imposed
for moving the value of the ${\mathrm{k^{th}}}$ input away from its standard value
by ${\rm{\sigma_k \;\xi_k}}$. The \chisq\ with pulls, which
includes the effects of all theoretical and systematic
uncertainties, is obtained by minimizing ${\rm{\chi^2(\xi_k)}}$
 with respect to all the pulls ${\rm{\xi_k}}$:
\begin{equation}  \label{eqn:chisqmin}
{\mathrm{ \chi^2_{pull} = Min_{\xi_k}\left[
\chi^2(\xi_k)\right] }}\,.
\end{equation}

\begin{table*}[t]
\tbl{Uncertainties for various quantities~\cite{raj2007}.}
{
\centering
\begin{tabular}{|c|c|} \hline
\textbf{Quantity} & \textbf{Value} \\ \hline
Flux normalization uncertainty & 20\% \\ \hline
Zenith angle dependence  uncertainty & 5\% \\ \hline
Cross section uncertainty& 10\% \\ \hline
Overall systematic uncertainty & 5\% \\ \hline
\multirow{2}{*}{Tilt uncertainty} & $ \Phi_\delta \approx \Phi_0(E) \left[1 + \delta
                                                                    \log\left(\frac{E}{E_0}\right)\right] $\\
                                  & with $ E_0 = 2 $ GeV, $ \sigma_\delta = 5\% $ (see, \eg, \cite{GonzalezGarcia:2004wg}) \\
\hline
\end{tabular} \label{tab:xsec-uncertainty}}
\end{table*}

In the calculation of $\chisq_{\text{pull}}$, we consider uncertainties in the flux, cross sections etc. (as in 
Table~ \ref{tab:xsec-uncertainty}), keeping the values of the oscillation parameters fixed while calculating $\rm{N^{ex}_{ij}}$ and $\rm{N^{th}_{ij}}$.
However, in general,
the values of the mass-squared difference $\Delta m_{31}^{2}$ and the mixing angles $\theta_{23}$ and $\theta_{13}$
can vary over a range corresponding to the actual measurements of these parameters. Holding them fixed at  particular values is equivalent to knowing the parameters to infinite precision, which is obviously unrealistic. To take into account the uncertainties in the actual measurement of the oscillation parameters, we define the marginalized \chisq\ as~\cite{raj2007} \\
\bea \label{eqn:chisqmarg}
{\mathrm{ \chi^2_{min}}}  &=& {\mathrm{ Min \left[ \chi^2 (\xi_k)
+ \left(\frac{|\Delta m_{31}^{2}|^{true} - |\Delta m_{31}^{2}|}{\sigma(|\Delta m_{31}^{2}|)}
\right)^2 \right.}}
\nonumber \\
&&{\mathrm{ \left. + \left( \frac{\sin^2 2\theta_{23}^{true} -
\sin^2 2\theta_{23}}{\sigma(\sin^2 2 \theta_{23})} \right)^2
+ \left( \frac{\sin^2 2\theta_{13}^{true} - \sin^2 2
\theta_{13}}{\sigma(\sin^2 2\theta_{13})}\right)^2 \right] }}\,.
\eea
The three terms added to $\chi^{2}(\xi_{k})$ are known as \emph{priors}. Now, for our \chisq\ analysis, we proceed as follows, {\it e.g.} for the case of the mass hierarchy.\\
\begin{itemize}

\item Our aim is to see at what statistical significance the \emph{wrong hierarchy} can be ruled out. Our procedure
gives the median sensitivity of the experiment in the frequentist approach~\cite{median}.

\item We simulate the number of events in 10~bins in the measured energy ${\rm{E_m}}$ and 
18~bins in the measured zenith angle ${\rm{\cos\theta_m}}$ for a set of \emph{true values} for the six
neutrino parameters: $\theta_{12}$, $\theta_{23}$, $\theta_{13}$,
$\Delta m_{21}^{2}$, $\Delta m_{31}^{2}$, \dcp\,, and for a \emph{true hierarchy}. The \emph{true values} are the current best fit values of the oscillation parameters and the true value of \dcp\ is assumed to be zero. This
is our \emph{experimental data} -- ${\rm N^{ex}_{ij}}$. Now we calculate the
\emph{theoretical expectation} in each bin -- ${\rm N^{th}_{ij}}$
assuming the \emph{wrong hierarchy}, and calculate the \chisq\ between these two datasets.

\item For the marginalization procedure, we allow  $\theta_{23}$,
$\theta_{23}$, $|\Delta m_{31}^{2}|$ and \dcp\ to vary within the following ranges:\\
$\theta_{23} \in [36^{\circ}, 54^{\circ}]$,\\ 
$\theta_{13} \in [5.5^{\circ}, 11^{\circ}]$,\\
$|\Delta m_{31}^{2}| \in [2.19, 2.62] \times 10^{-3}$ eV$^{2}$,\\
$\dcp \in [-\pi, \pi]$.

\item In computing ${\rm{\chi^2_{min}}}$, we
add the priors for the neutrino parameters
which assigns a penalty for moving away from the true value. 
During marginalization, as the value of an oscillation parameters shifts further from its true value, Eq.~(\ref{eqn:chisqmarg}) suggests that the corresponding prior will be larger resulting in a higher \chisq\ value.

\item Finally, after adding the priors, we determine $\mathrm{\chisq_{pull}}$ (see Eq.~\ref{eqn:chisqmin}). This is a multi-dimensional parameter space minimization of the function $\chisq(\alpha, \beta, \dots)$, where $\alpha$, $\beta, \ldots$ are the parameters over which marginalization is required. For the purpose of this multi-minimization, we have used the NLopt library~\cite{nlopt}. We do the minimization first over the entire multi-dimensional parameter space to locate the global minimum approximately, and then use the parameters corresponding to this as a guess to carry out a local minimum search to locate the minimized \chisq\ within the parameter space accurately.
We carry out this minimization routine using a simplex algorithm described in Ref.~\cite{NelderMead} , and implemented within the NLopt library.

\end{itemize}

\section{DUNE fluxes and atmospheric neutrino events}
The fluxes and charged current cross sections used in our analysis are shown in Fig.~\ref{fig:fluxsec}. These are similar
to those used by the DUNE collaboration.

\begin{figure*}[tb]
	\centering
	\begin{subfigure}
		\centering
		\includegraphics[scale = 0.39]{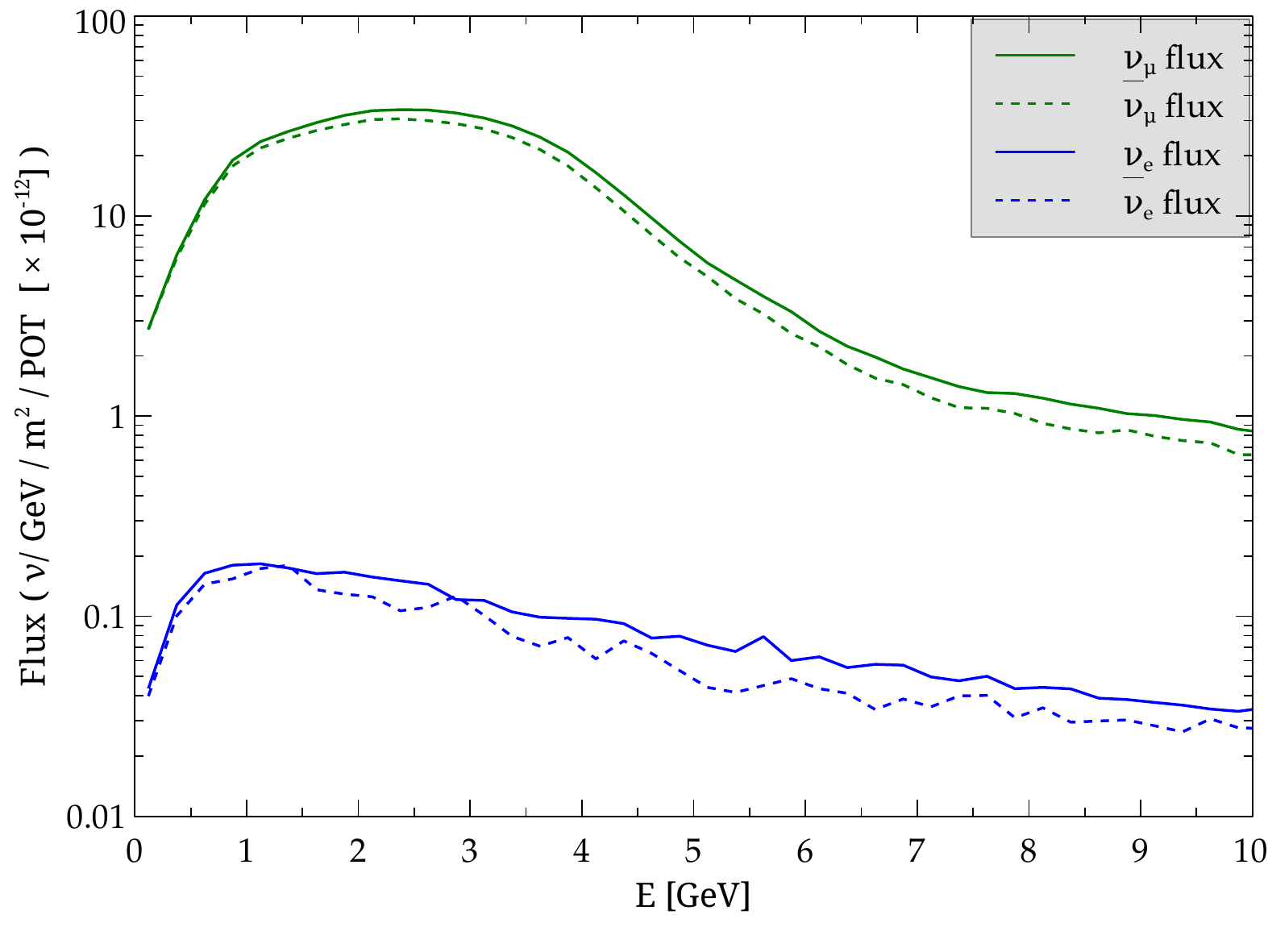}
		\end{subfigure}
	\begin{subfigure}
		\centering
		\includegraphics[scale = 0.39]{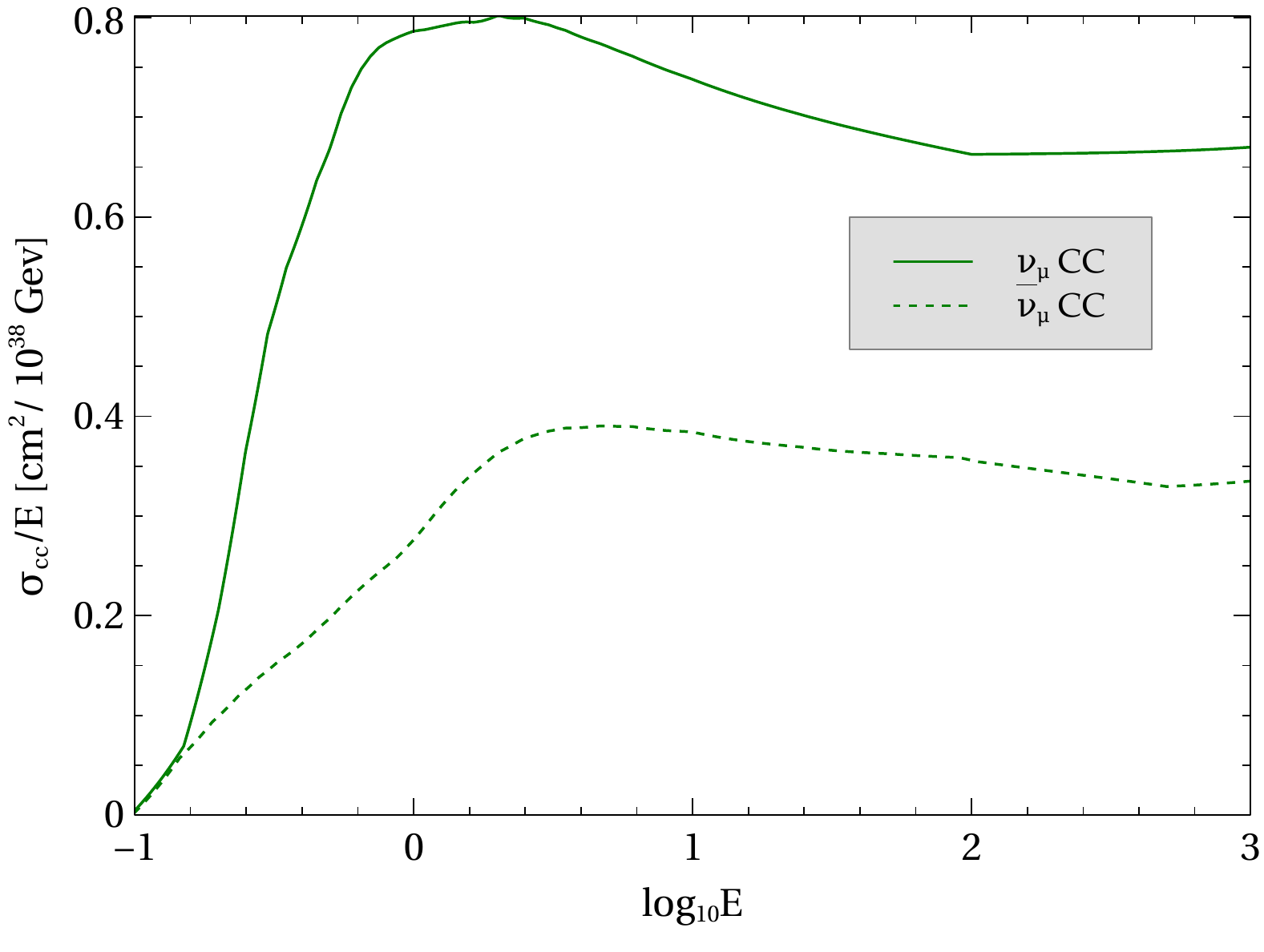}
		\end{subfigure}
	\caption{The neutrino and antineutrino fluxes are shown in the left panel. The right panel shows the $\nu$ and $\bar{\nu}$ charged current cross sections.}
	\label{fig:fluxsec}
\end{figure*}
%
%%\section{Events}
In Fig.~\ref{fig:atmos_event}, we show the number of \numu\ and \nue\ atmospheric events with and without oscillations for an exposure of 350 kt-yr.

\begin{figure*}[tb]
	\centering
	\begin{subfigure}
		\centering
		\includegraphics[scale = 0.39]{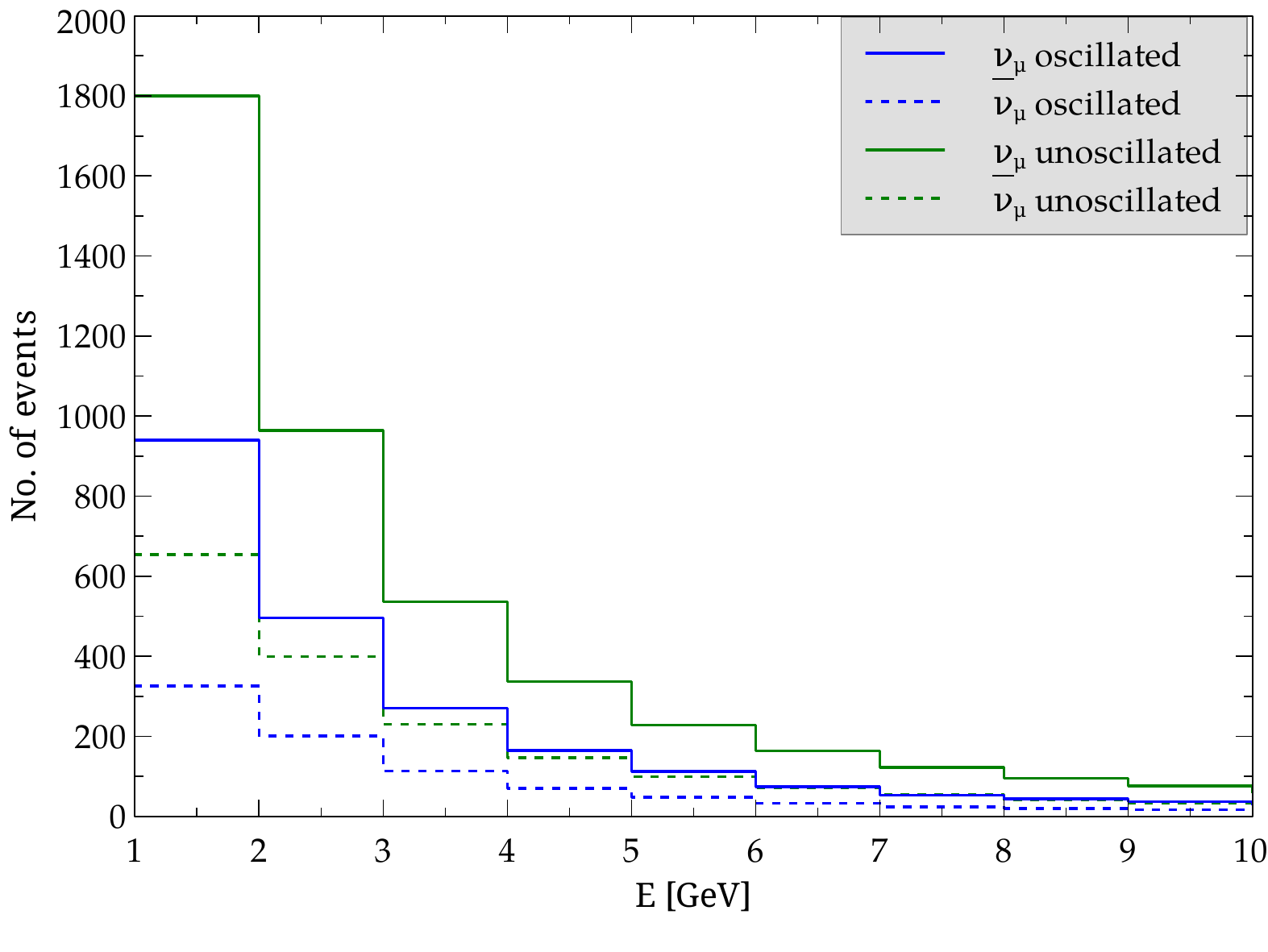}
		\end{subfigure}
	\begin{subfigure}
		\centering
		\includegraphics[scale = 0.39]{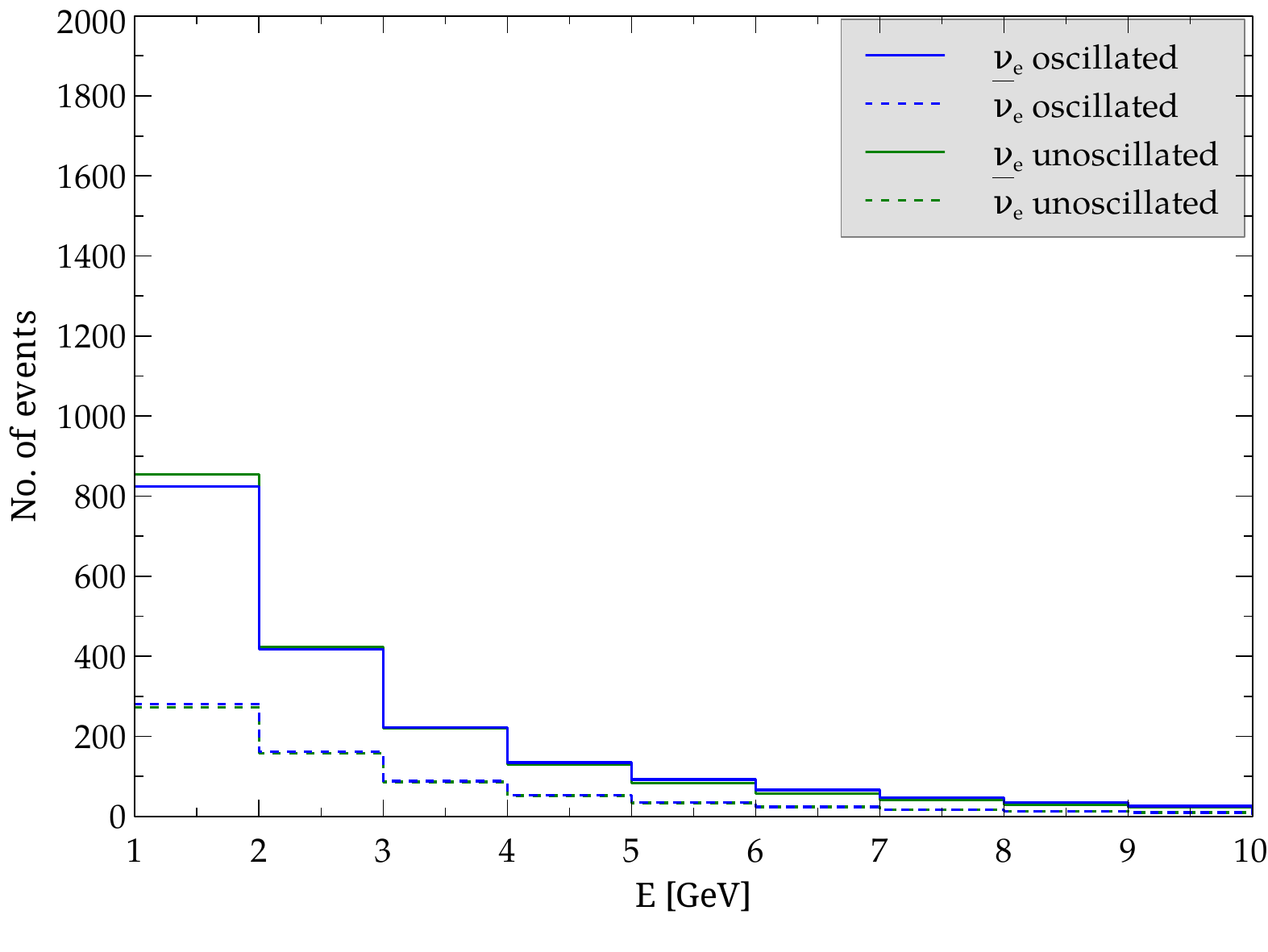}
		\end{subfigure}
	\caption{\numu\ (left panel) and \nue\ (right panel) atmospheric events for a 350 kt-yr \liar\ FD.}
	\label{fig:atmos_event}
\end{figure*}

\end{document}